\documentclass[twocolumn,showpacs,preprintnumbers,amsmath,amssymb,superscriptaddress]{revtex4}

\usepackage{graphicx}
\usepackage{dcolumn}
\usepackage{bm}
\usepackage{tabularx}
\usepackage{ulem} 
\newcolumntype{M}{>{\centering\arraybackslash}m{1.85cm}}
\usepackage[export]{adjustbox}
\usepackage{float}
\usepackage{color}   
\usepackage{hyperref}
\hypersetup{
    colorlinks=true, 
    linktoc=all,     
    linkcolor=blue,  
}

\makeatletter
\newcommand{\colorcaption}[2][]{%
  \begingroup%
  \renewcommand{\@caption@fignum@sep}{ (Color online). }%
  \caption[#1]{#2}%
  \endgroup%
}
\makeatother
\begin{document}

\title{\textit{Ab-initio} no-core shell model study of $^{10-14}$B isotopes with realistic \textit{NN} interactions}

\author{ Priyanka Choudhary\footnote{pchoudhary@ph.iitr.ac.in}, Praveen C. Srivastava\footnote{Corresponding author: praveen.srivastava@ph.iitr.ac.in}}
\address{Department of Physics, Indian Institute of Technology Roorkee, Roorkee 247667, India}
\author{ Petr Navr\'atil\footnote{navratil@triumf.ca}}
\address{TRIUMF, 4004 Wesbrook Mall, Vancouver, British Columbia V6T 2A3, Canada}

\date{\hfill \today}


\begin{abstract}
	
We report a comprehensive study of $^{10-14}$B isotopes within the \textit{ab-initio} no-core shell model (NCSM) using realistic nucleon-nucleon (\textit{NN}) interactions.
In particular, we have applied the inside non-local outside Yukawa (INOY) interaction to study energy spectra, electromagnetic properties and point-proton radii of the boron isotopes. The NCSM results with the charge-dependent Bonn 2000 (CDB2K), the chiral next-to-next-to-next-to-leading order (N$^3$LO) and optimized next-to-next-to-leading order (N$^2$LO$_{opt}$) interactions are also reported.
We have reached basis sizes up to $N_{\mbox{max}}$ = 10 for $^{10}$B, $N_{\mbox{max}}$ = 8 for $^{11,12,13}$B and $N_{\mbox{max}}$ = 6 for $^{14}$B with m-scheme dimensions up to 1.7 billion. We also compare the NCSM calculations with the phenomenological YSOX interaction using the shell model to test the predictive power of the \textit{ab-initio} nuclear theory. Overall, our NCSM results are consistent with the available experimental data. 
The experimental ground state spin $3^{+}$ of $^{10}$B has been reproduced using the INOY \textit{NN} interaction. Typically, the 3\textit{N} interaction is required to correctly reproduce the aforementioned state.

\end{abstract}


\maketitle
\newpage
\section{Introduction}

In nuclear physics, our focus is to describe the nuclear structure including the exotic behaviour of atomic nuclei throughout the nuclear chart. Conventional shell model \cite{SM,p,sd,fp,O.Sorlin,T.Otsuka}, where interactions are assumed to exist only among the valence nucleons in a particular model space is unable to determine the drip line  \cite{dripline1,dripline2},  cluster  \cite{cluster} and halo \cite{halo} structures.
The study of interactions derived  from first principles has been a challenging area of research over the past decades. These fundamental interactions are determined from either meson-exchange theory or Quantum chromodynamics (QCD) \cite{QCD}. QCD is non-perturbative in low-energy regime which makes analytic  solutions difficult. This difficulty is overcome by chiral effective field theory ($\chi$EFT) \cite{RMP,EFT1,EFT2,EFT3}. Chiral perturbation theory ($\chi$PT) \cite{PT} within $\chi$EFT provides a connection between QCD and the hadronic system.

Progress has been made in the development of  different many-body modern \textit{ab-initio} approaches \cite{ab-initio,HHMS2015,SHKH2019}, one of them being the NCSM \cite{Phys.Rev.C502841(1994),Phys.Rev.C542986(1996),Phys.Rev.C573119(1998),PRL84,1999,PRC62,2000,2009,MVS2009,nocore_review,stetcu1,stetcu2}. \textit{Ab-initio} methods are more fundamental compared to the nuclear shell model.
The aim of this paper is to explain the nuclear structure of boron isotopes with realistic \textit{NN} interactions as the only input. The well-bound stable $^{10}$B have posed a challenge to the microscopic nuclear theory in particular concerning the reproduction of its ground-state spin~\cite{av8'nocore}. The boron isotopes have been investigated in the past using the shell model \cite{monopole,TOSM}.
Shell model Hamiltonian constructed from a monopole-based universal interaction (V$_{MU}$) in full $psd$ model space
including $(0{-}3)\hbar\Omega$ excitations has been used for a systematic study of boron isotopes \cite{monopole}. This phenomenological effective 
interaction is obtained by fitting experimental data, thus, it at least partly includes three-body effects. So it is able to 
reproduce spin of the ground state (g.s.) of $^{10}$B. This V$_{MU}$ based Hamiltonian, however, fails to describe the drip line nucleus $^{19}$B.
Tensor-optimized shell model (TOSM) \cite{TOSM} has been applied to study $^{10}$B using effective bare nucleon-nucleon (\textit{NN}) interaction Argonne V8$^{'}$ (AV8$^{'}$) \cite{av8'}. The g.s. obtained with AV8$^{'}$ interaction is $1^{+}$, which, in experiment, is the first excited state of $^{10}$B. AV8$^{'}_{eff}$ interaction,
which is a modification of tensor and spin-orbit forces of AV8$^{'}$ interaction, gives correct g.s. spin and low-lying spectra, indicating that the tensor forces affect the level ordering. TOSM with Minnesota (MN) effective interaction \cite{MN} without tensor force also
gives correct g.s. spin but  a smaller g.s. radius compared to the experimental result, which affects the nuclear saturation property, thus providing the small level density. 

In Refs. \cite{Nmax8,10B_PRL,Nmax10}, the structure of $^{10}$B was studied within the NCSM, using accurate charge dependent \textit{NN} potentials up to the $4^{th}$ order of $\chi$PT in basis spaces ($N_{\mbox{max}}$) of up to $10\hbar\Omega$. Using the \textit{NN} interactions alone led to an incorrect g.s. of $^{10}$B. By including the chiral three-nucleon interaction (3\textit{N}), the g.s. was correctly reproduced as $3^{+}$ \cite{10B_PRL,Nmax10}.
The \textit{ab-initio} NCSM study of $^{10}$B 
with the chiral N$^{2}$LO (next-to-next-to-leading order) \textit{NN} interaction \cite{EKM2015}
including three-body forces has been done in Ref. \cite{N2LO}, where it was shown that the g.s. energy and spin depends on the chiral order. To correctly reproduce the $3^+$ as an experimental g.s., 3\textit{N} force with the N$^{2}$LO \textit{NN} interaction is needed. 	
In Ref. \cite{N2LOopt}, N$^{2}$LO$_{opt}$ interaction was employed in the NCSM calculation for $^{10}$B up to $N_{\mbox{max}}$ = 10 (10$\hbar\Omega$) to calculate ground and low-lying excited states. This study reported $1^+$ as a g.s. instead of $3^+$. 
Realistic shell model calculations including  contributions of a chiral three-body force 
[N$^3$LO \textit{NN} + N$^2$LO 3\textit{N} potential] for $^{10}$B is reported in Ref. \cite{MBPT}. 
These results are consistent with the  NCSM results with  the same interaction.
The NCSM with CDB2K potential ($N_{\mbox{max}}$ = 8) and AV8' ($N_{\mbox{max}}$ = 6) predict $1^{+}$ as g.s. of $^{10}$B \cite{CDB2Knocore,av8'nocore}. Green's function Monte Carlo (GFMC) approach with AV8' and AV18 has also been employed to investigate the g.s. of $^{10}$B \cite{GFMC} and similarly predicts the $1^+$ as the ground state with these \textit{NN} forces.

In Ref. \cite{PLB}, the Daejeon16 and JISP16 (J-matrix inverse scattering potential) \textit{NN} interactions  were applied to $p$-shell nuclei. For $^{10}$B, 
excitation energies of $1^{+}$ state with respect to $3^{+}$ state of 0.5(1) MeV and 0.9(2.4) MeV were reported with Daejeon16 and JISP16 \textit{NN} interactions, respectively. This means both these \textit{NN} interactions reproduce correct g.s. without adding 3\textit{N} forces but the ordering could not be confirmed on the account of uncertainty in the energy result obtained from JISP16 interaction.


In recent years, several experimental techniques  have  been used to measure nuclear charge radius for neutron-rich nuclei towards the drip line \cite{nature_Ca}. These then serve as a test of the predictive power of \textit{ab-initio} calculation. 
Charge radii inform us about the breakdown of the conventional shell gaps and the evolution of new shell gaps.
One of the  reasons behind  the disappearance of the shell gap is the presence of the halo structure.
Tanihata \textit{et al.} \cite{RMSradii} have measured  interaction  cross-sections ($\sigma_{I}$) for $^{8,12-15}$B using radioactive nuclear beams at the Lawrence Berkeley Laboratory. In this  experiment,  the interaction nuclear radii  and the effective root-mean-square (rms) radii of nucleon distribution have been deduced from $\sigma_{I}$. 
Point-proton radii of $^{12-17}$B are also measured from  the charge-changing cross-section ($\sigma_{cc}$) at GSI, Darmstadt \cite{N=14}. 
Further, the proton radii  were extracted from a finite-range Glauber model analysis of the $\sigma_{cc}$. The measurement  shows the existence of a thick neutron surface in $^{17}$B \cite{chargechanging}.
A recent  experiment on nitrogen chain  establishes the neutron skin and signature of the $N=14$ shell gap by measuring  proton-radii of $^{17-22}$N 
isotopes.

In the present work, we perform systematic NCSM calculations for $^{10-14}$B isotopes using INOY~\cite{INOY}, N$^{3}$LO~\cite{Machleidt}, CDB2K~\cite{CDBonn} and N$^{2}$LO$_{opt}$~\cite{N2LOopt} \textit{NN} interactions. For the first time, we report NCSM structure results with the INOY interaction for these isotopes. We have reached basis sizes up to $N_{\mbox{max}}$ = 10 for $^{10}$B, $N_{\mbox{max}}$ = 8 for $^{11,12,13}$B and $N_{\mbox{max}}$ = 6 for $^{14}$B with m-scheme dimensions up to 1.7 billion.  Apart from energy spectra, we have also calculated electromagnetic properties and point-proton radii. 
In addition, we compare shell model results of energy levels and nuclear observables obtained with the YSOX interaction~\cite{monopole}  with present \textit{ab-initio} results.

The paper is organized as follows: In section II, we describe the NCSM formalism. 
In section III, we briefly review the \textit{NN} interactions used in our calculations. 
We present the NCSM results of the energy spectra 
and compare them to those obtained with the shell model YSOX interaction in section IV. 
In section V,  electromagnetic properties of $^{10-14}$B 
are reported. 
In section VI,  we discuss point-proton radii of $^{10-14}$B. 
Finally, we summarize the paper in section VII.

\section{No-core shell model formalism}

In NCSM \cite{2009,nocore_review}, all nucleons are treated as active, which  means there is no assumption of an inert core, unlike in standard shell model.
The nucleus is described as a system of $A$ non-relativistic nucleons which  interact
by realistic \textit{NN} or \textit{NN} + 3\textit{N} interactions.

In the present work, we have considered only realistic \textit{NN} interactions between the nucleons.
The Hamiltonian for the $A$ nucleon system is then given by
\begin{equation}\label{hamil}
H_{A} = T_{rel} + V = \frac{1}{A}\sum_{i<j}^{A} \frac{{(\vec p_i - \vec p_j)}^2}{2m} + \sum_{i<j}^{A} V^{NN}_{ij},
\end{equation}
where T$_{rel}$ is the relative kinetic energy, $m$ is the mass of nucleon and V$^{NN}_{ij}$ is the realistic
\textit{NN} interaction that contains both nuclear and electromagnetic (Coulomb) parts. 

In the NCSM, translational invariance as well as angular momentum and parity of the nuclear system are conserved. The many-body wave function is cast into an expansion over a complete set of antisymmetric $A$-nucleon harmonic oscillator (HO) basis states containing up to $N_{\rm max}$\,- HO excitations above the lowest possible configuration.

We use a truncated HO basis while the realistic \textit{NN} interactions act in the full space. Unless the potential is soft like, e.g., the N$^{2}$LO$_{opt}$, we need to derive an effective interaction to facilitate the convergence. Two renormalization methods based on similarity transformations have been applied in the NCSM, the Okubo-Lee-Suzuki (OLS) scheme \cite{Prog.Theor.Phys.12,Prog.Theor.Phys.,Prog.Theor.Phys.68,Prog.Theor.Phys.92} and more recently the Similarity Renormalization Group (SRG)~\cite{Bogner2007}. The latter has the advantage in being more systematic and in the fact that renormalized potentials are phaseshift equivalent. The three-body induced terms, however, cannot be neglected. Those, in turn, are difficult to converge for potentials that generate strong short-range correlations, such as the CDB2K~\cite{Jurgenson2011}. The OLS method is applied directly in the HO basis and results in an $A$- and $N_{\rm max}$-dependent effective interaction, i.e, the calculation is not variational. The three-body induced terms are less important. It has been observed that the method works particularly well for the INOY interaction~\cite{Forssen2005,Caurier2006,Forssen2009,Forssen2013}. Consequently, in this work we apply the OLS method for the INOY, CDB2K and, for a consitent comparison also for the N$^3$LO \textit{NN} interaction. For the latter, the SRG method is, however, more appropriate~\cite{Jurgenson2011,Jurgenson2013}. The softer N$^{2}$LO$_{opt}$ \textit{NN} interaction is not renormalized.

To facilitate the derivation of the OLS effective interaction, we add centre-of-mass (c.m.) HO Hamiltonian to equation (\ref{hamil}) which makes the Hamiltonian dependent on the HO frequency.
\begin{equation*}
	H_{\mbox{c.m.}} = T_{\mbox{c.m.}} +U_{\mbox{c.m.}},\\
\end{equation*}
where
\begin{equation*}
	 U_{\mbox{c.m.}} = \frac{1}{2}Am{\Omega}^2{\vec R}^2;
\end{equation*}
\begin{equation*}
\vec R = \frac{1}{A} \sum _{i=1}^{A} \vec r_i.
\end{equation*}
The intrinsic properties of the system are not affected by the addition of HO c.m. Hamiltonian due to translational invariance of the Hamiltonian (\ref{hamil}). 

Thus, we obtain a modified Hamiltonian:
\begin{equation}\label{(2)}
H^{\Omega}_{A} = H_{A} + H_{\mbox{c.m.}} = \sum _{I=1}^{A} h_{i} + \sum _{i<j}^{A}V^{\Omega,A}_{ij} \\
\nonumber
\end{equation}
\begin{equation}\label{(2)}
= \sum_{i<j}^{A} \left[\frac{{\vec p_i}^2}{2m}+\frac{1}{2}m {\Omega}^2 {\vec r_i}^2 \right]
+ \sum_{i<j}^{A} \left[V_{ij}^{NN}-\frac{m {\Omega}^2}{2A} {(\vec r_i - \vec r_j)}^2 \right].
\end{equation}
We divide the A nucleon large HO basis space into two parts: one is the finite active 
space ($P$) which contains all states up to $N_{\mbox{max}}$, 
and, the other is the excluded space ($Q=1-P$). NCSM calculations are performed in the truncated $P$ space. The two-body OLS effective is derived by applying the Hamiltonian (\ref{(2)}) to two nucleons and performing the unitary transformation in the HO basis~\cite{2009,nocore_review}. Eventually, the second term in the brackets in (\ref{(2)}) is replaced by the effective interaction.

Finally, we subtract the c.m. Hamiltonian $H_{\mbox{c.m.}}$ and include the Lawson projection term \cite{Lawson} to shift the spurious c.m. excitations.
\begin{multline}\label{(3)}
	H_{A,eff}^{\Omega} = P\left\{ \sum _{i<j}^{A} \left[ \frac{{(\vec p_i - \vec p_j)}^2}{2mA} + \frac{m {\Omega}^2}{2A} {(\vec r_i - \vec r_j)}^2 \right]\right. \\
	\left.+\sum_{i<j}^{A} \left[ V^{NN}_{ij} - \frac{m {\Omega}^2}{2A}{(\vec r_i - \vec r_j)}^2\right]_{\rm eff}+\beta \Bigg(H_{\mbox{c.m.}} - \frac{3}{2}\hbar\Omega\Bigg)\right\}P.\\
	\quad
\end{multline}

An extension of the NCSM that provides a unified description of both bound and unbound states is the no-core shell model with continuum (NCSMC) approach~\cite{NCSMC}. It has been successfully applied, e.g., to explain the parity inversion phenomenon in $^{11}$Be~\cite{11Be}. It has not been applied to boron isotopes yet although NCSMC calculations for $^{10,11}$B are now in progress.


\section{Realistic \textit{NN} and shell model interactions}

In the present work, apart from the INOY interaction \cite{INOY,nonlocal,Doleschall}, we also report results with the CDB2K \cite{offshell,Phys.Rep.,Adv.Nucl.Phys.,CDBonn}, N$^{3}$LO \cite{QCD,Machleidt} and N$^{2}$LO$_{opt}$ \cite{optimizedn2lo,N2LOopt} interactions.


The Inside Non-Local Outside Yukawa (INOY) interaction \cite{INOY,nonlocal,Doleschall} has a local character (Yukawa tail) at long distances ($r\geq3$ fm) and a non-local one at short distances ($r\, {<}\,3$ fm), where  the non-local part is due to the internal structure of  the nucleon. As it is constructed in coordinate space, the range of locality and non-locality is explicitly controllable. This interaction has the form:  
\begin{align}
	V^{full}_{ll'}(r,r') & = W_{ll'}(r,r')+ \delta (r-r') F^{cut}_{ll'}(r) V^{Yukawa}_{ll'}(r),\notag \\ & 
\end{align}
where, the cut-off function is defined as: 
\[ F^{cut}_{ll'}(r) =
\begin{cases}
1- e^{-[\alpha_{ll'}(r-R_{ll'})]^{2}} & for ~r\geq R_{ll'} ,\\
0 & for ~r\leq R_{ll'} , \\
\end{cases}
\]
\\
and $W_{ll^{'}}(r,r^{'})$ and $V_{ll^{'}}^{Yukawa}(r)$ are the non-local part and the Yukawa tail (the same as in AV18 potential \cite{av18}), respectively. The parameters $\alpha_{ll'}$ and $R_{ll'}$ have the values 1.0 fm$^{-1}$ and 2.0 fm, respectively.
Because of the non-local character in the INOY interaction, three-body force effects are in part absorbed by nonlocal terms, e.g., it produces correct binding energy of the three-nucleon system ($^3$H and $^3$He) without adding three-body 
forces explicitly.


The Charge-Dependent Bonn 2000 potential (CDB2K) is a meson exchange based potential \cite{offshell,Phys.Rep.,Adv.Nucl.Phys.,CDBonn}.  It includes all the mesons with masses below the nucleon mass, i.e. $\pi^{\pm,0}$, $\eta$, $\rho^{\pm,0}$ and $\omega$ as an exchange particle between nucleons. The $\eta$ has 
a vanishing coupling constant  and as such, can be ignored. This potential also includes two scalar-isoscalar $\sigma$ (or $\epsilon$) bosons.
 Charge dependence of nuclear forces, which is investigated by the Bonn full model based on charge independence breaking (difference between proton-proton/neutron-neutron and proton-neutron interaction; pion mass splitting) and charge symmetry breaking (difference between proton-proton and neutron-neutron interaction; nucleon mass splitting) in all partial waves with $J\leq4$, is also reproduced. 
The potential is represented in terms of the one-boson-exchange (OBE) covariant Feynman amplitudes. The off-shell behavior of the potential, which plays an important role in nuclear structure calculations, is affected by imposing locality on the Feynman amplitudes. 
So, non-local Feynman amplitudes are used in the CDB2K potential.
This momentum-space dependent potential fits proton-proton data with $\chi^{2}$ per
datum of 1.01 and the neutron-proton data with $\chi^{2}$/datum $=$ 1.02 below 350 MeV, where $\chi^{2}$ is the square of theoretical error over the experimental error.

Chiral perturbation theory is a perturbative expansion in $Q/\Lambda_{\chi}$, where $Q\ll\Lambda_{\chi}\approx$ 1 GeV. Entem and Machleidt constructed the \textit{NN} potential \cite{QCD,Machleidt} at fourth order (next-to-next-to-next-to-leading order; N$^{3}$LO) of $\chi$PT in the momentum-space.
In $\chi$PT, two class of contributions determine the \textit{NN} amplitude:  Contact terms and pion-exchane diagrams. The N$^{3}$LO interaction contains 24 contact terms, whose parameters contribute  to  the fit of partial waves of \textit{NN} scattering with angular momentum $L\leq2$. Charge dependence is also included up to next-to-leading order of the isospin-violation scheme. The N$^{3}$LO has two charge-dependent contacts. Thus,  the total number of contact terms 
is 26. The N$^{3}$LO has one pion-exchange (OPE) as well as two pion-exchange (TPE) contributions. Contributions of three pion exchange in the N$^{3}$LO, however, are negligible. OPE and TPE depend on the axial-vector coupling constant $g_{A}$ (1.29), the pion decay constant $f_{\pi}$ (92.4 MeV) and eight low-energy constants (LEC). Three of them ($c_2$, $c_3$ and $c_4$) are varied in the fitting process and other are fixed. All constants are determined from the \textit{NN} data. 
With a total of 29 parameters, the N$^{3}$LO yields $\chi^{2}$/datum $\approx$ 1 up to 290 MeV for the fit of neutron-proton data. The accuracy in the  reproduction of \textit{NN} data for this order is comparable to the high-precision phenomenological AV18 potential \cite{av18}.

The N$^{2}$LO$_{opt}$ \cite{N2LOopt,optimizedn2lo} is a softer interaction and as such, the OLS or SRG renormalization is not needed. This interaction was derived from $\chi$EFT at the N$^{2}$LO order. For the optimization of the LECs, Practical Optimization Using No Derivatives algorithm (POUNDERs) was used. In particular, the optimisation is performed for the pion-nucleon ($\pi$N) couplings ($c_{1}$, $c_{3}$, $c_{4}$) and 11 partial wave contact parameters $C$ and $\tilde{C}$. 
The N$^{2}$LO$_{opt}$ interaction reproduces reasonably well experimental binding energies and radii of $A$ = 3, 4 nuclei. 

For comparison, we have also performed shell model calculations with the phenomenological YSOX interaction \cite{monopole} developed by  the Tokyo group. In the YSOX interaction, $^4$He is assumed as a core and interactions take place in the $psd$ valence space. Single-particle energies are $e_{p_{3/2}}$ = 1.05 MeV, $e_{p_{1/2}}$ = 5.30 MeV, $e_{d_{5/2}}$ = 8.01 MeV, $e_{s_{1/2}}$ = 2.11 MeV and $e_{d_{3/2}}$ = 10.11 MeV. There are 516 two-body matrix elements (TBMEs) in this interaction.

NCSM calculations presented in this paper have been performed with the pAntoine code \cite{pAntoine1,pAntoine11,pAntoine2}. We have used KSHELL code  \cite{KSHELL2019}  for the shell model calculation with the YSOX interaction \cite{monopole}. Recently, we have reported NCSM results for N, O and F isotopes in Refs. \cite{arch1,arch2} performed in an analogous way.

\section{Results and Discussions}

\begin{figure*}
	\includegraphics[width=8cm]{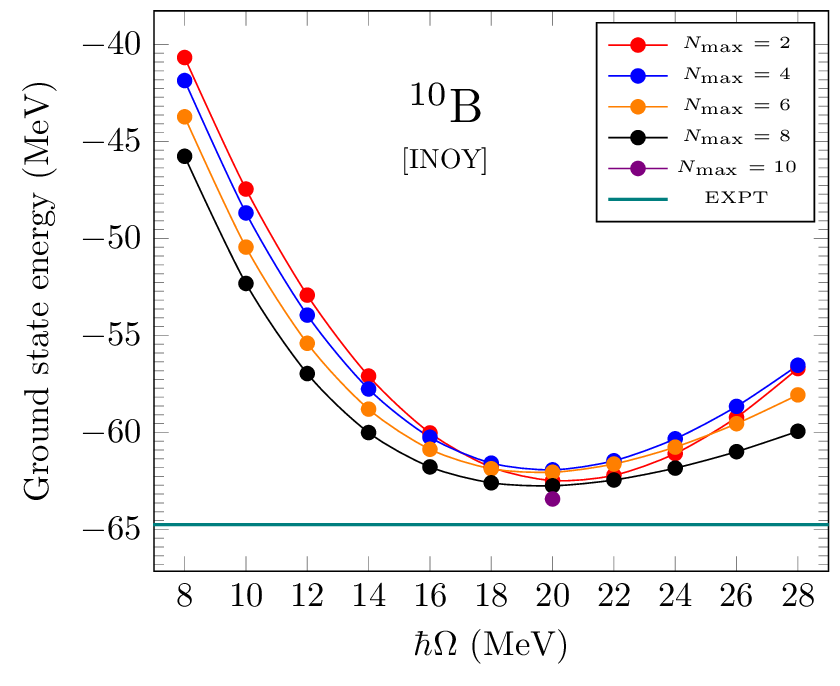}
	\includegraphics[width=8cm]{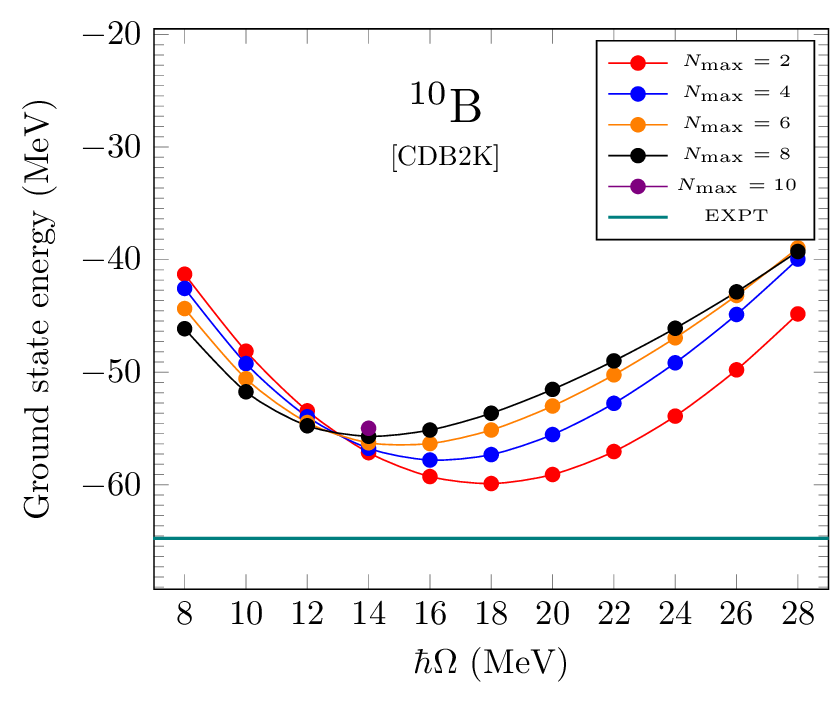}
	\includegraphics[width=8cm]{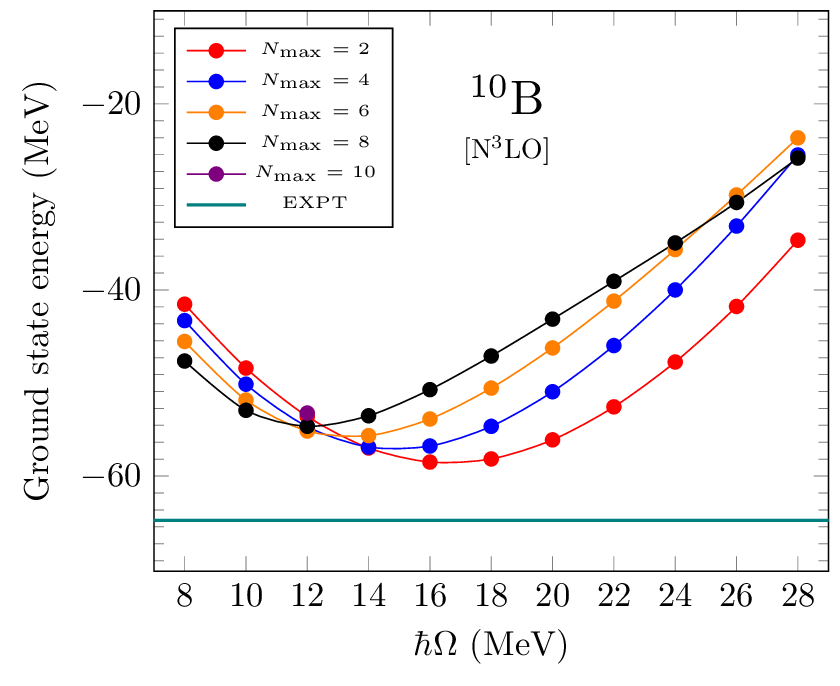}
	\includegraphics[width=8cm]{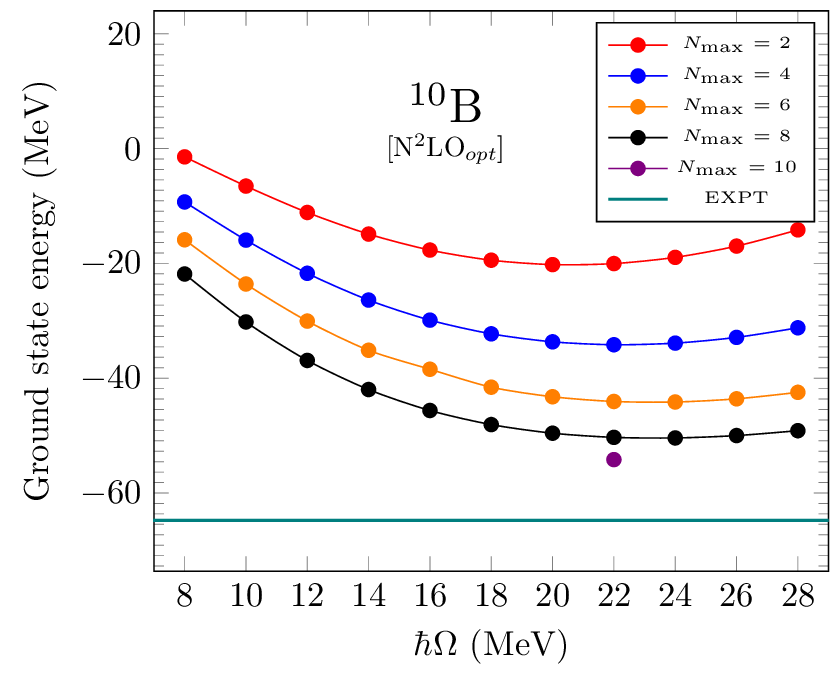}
	\caption{Ground state energy of $^{10}$B as a function of HO frequency for $N_{\mbox{max}}$ = 2 to 10 with the INOY, CDB2K, N$^3$LO and N$^{2}$LO$_{opt}$ interactions. Experimental g.s. energy is shown by  the horizontal line.}
	\label{basis}
\end{figure*}

The dimensions corresponding to different $N_{\mbox{max}}$ for boron isotopes are shown in Table \ref{tab:my_label}. We can see that they increase rapidly with $N_{\mbox{max}}$ and the mass number.
In the present work, we were able to perform NCSM calculations up to $N_{\mbox{max}}$ = 10 for $^{10}$B, $N_{\mbox{max}}$ = 8
for $^{11,12,13}$B and $N_{\mbox{max}}$ = 6 for $^{14}$B.  First, we investigate the dependence on the HO frequency ($\hbar\Omega$) for various $N_{\mbox{max}}$ bases, typically up to the next to the largest accesible for computational reasons.
The optimal HO frequency used to calculate the entire energy spectrum is found from the g.s. energy minimum in the largest $N_{\mbox{max}}$ space.
Fig. \ref{basis} shows variation of g.s. energy of $^{10}$B for different basis spaces as a function of HO frequencies for the four interactions that we employ. Overall, we observe a decrease of the g.s. energy dependece on the frequency at higher $N_{\mbox{max}}$ as expected. Let us re-iterate that the N$^{2}$LO$_{opt}$ calculations are variational while those with the OLS renormalized interactions are not.

\begin{figure*}
	\includegraphics[width=7.5cm]{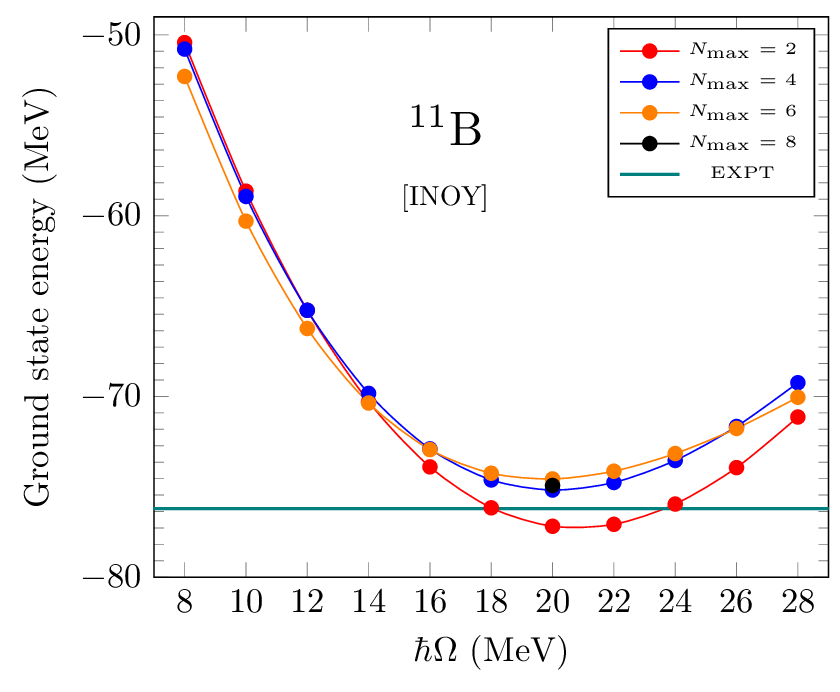}
        \includegraphics[width=7.5cm]{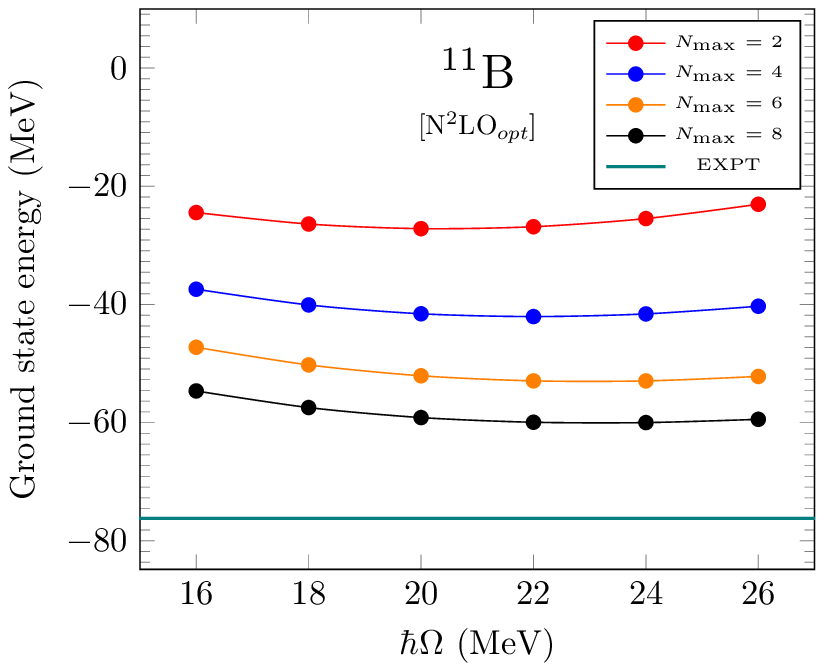}
        \includegraphics[width=7.5cm]{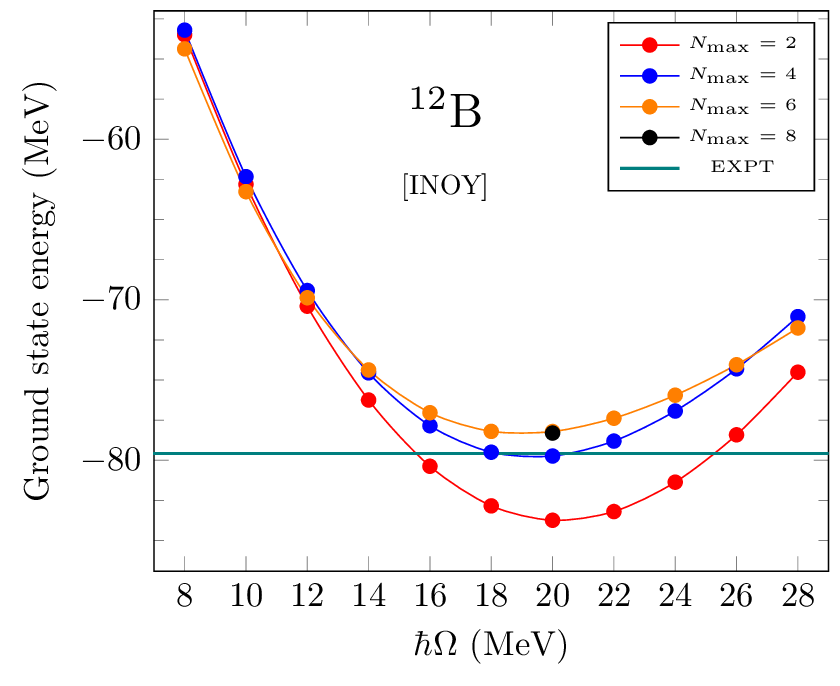}
         \includegraphics[width=7.5cm]{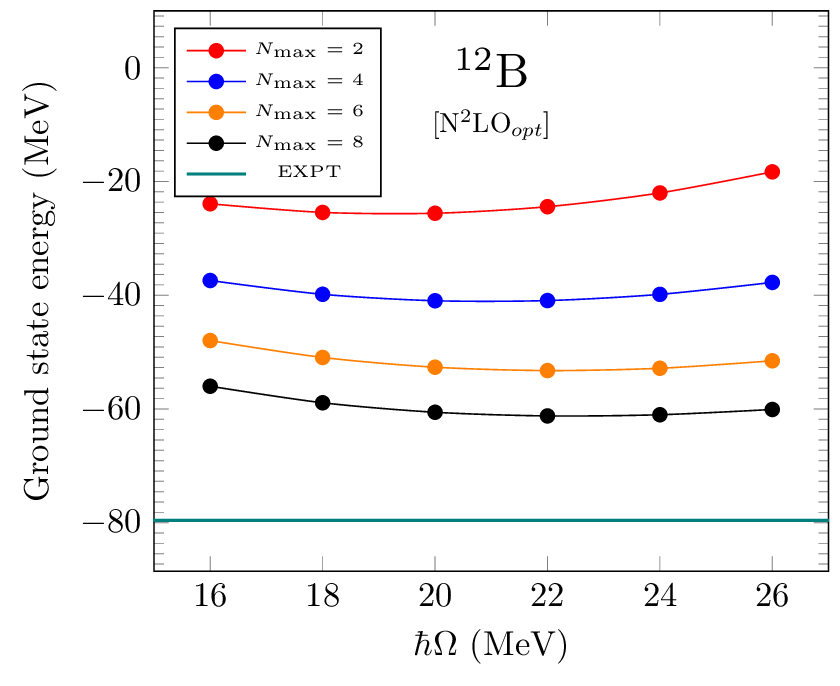}
         \includegraphics[width=7.5cm]{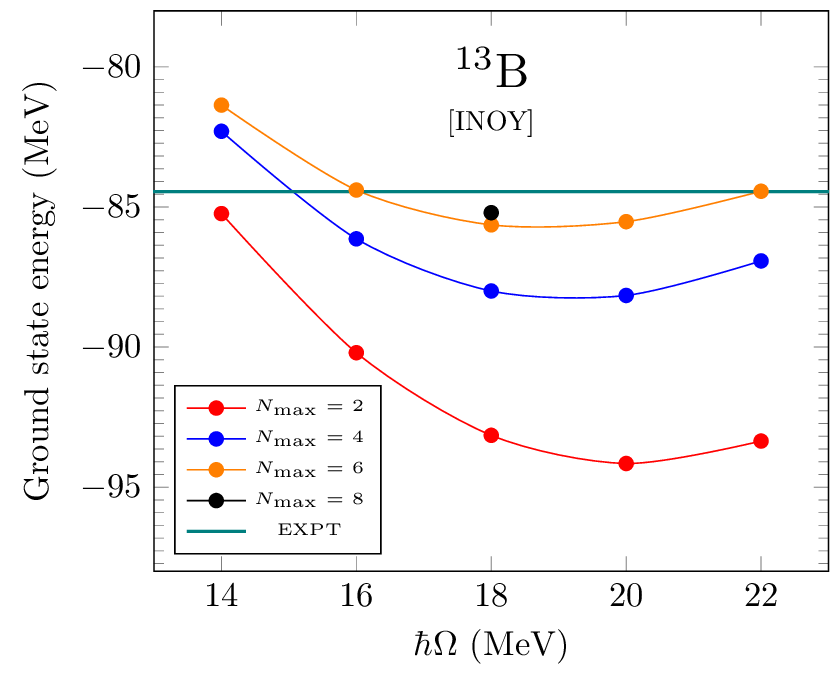}
         \includegraphics[width=7.5cm]{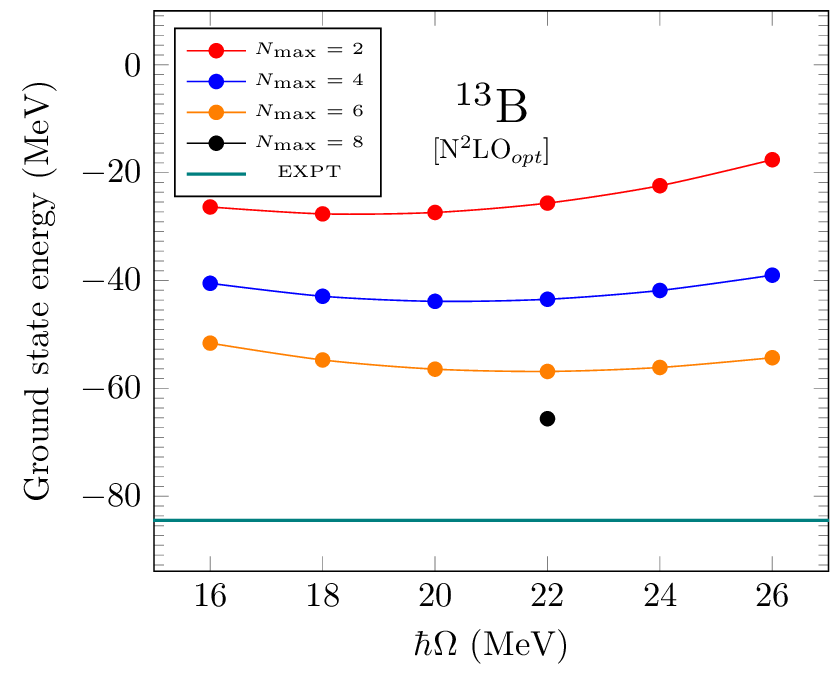}
         \includegraphics[width=7.5cm]{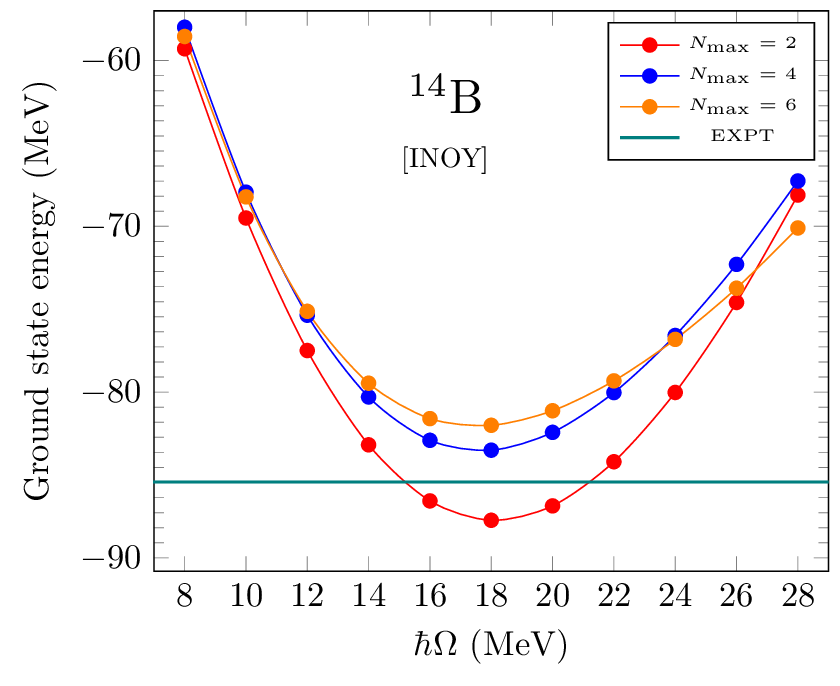}
          \includegraphics[width=7.5cm]{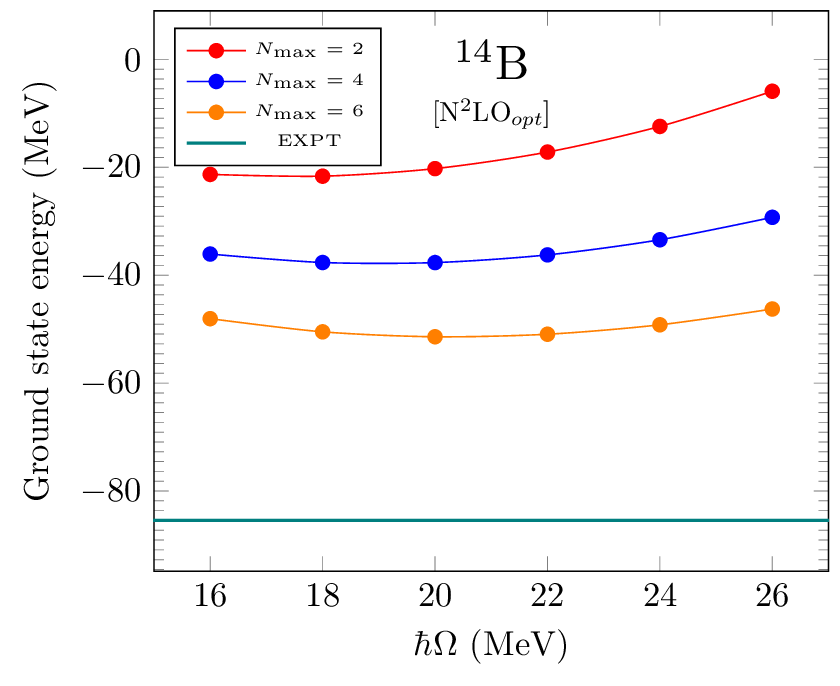}        
	\caption{Ground state energy of $^{11,12,13,14}$B as a function of HO frequency for different $N_{max}$ with the INOY and N$^{2}$LO$_{opt}$ interactions. }
	\label{basis1}
\end{figure*}

\begin{table}[ht]
	\centering
	\caption{Dimensions in m-scheme for boron isotopes corresponding to different $N_{max}$. The dimensions  up to which we have reached are shown in blue. }
	\label{tab:my_label}
	\begin{adjustbox}{width=0.48\textwidth}
		\begin{tabular}{cccccc}
			\hline
			\hline \vspace{-2.8mm}\\
			\hspace{-0.1cm}$N_{\mbox{max}}$  &  $^{10}$B  &  $^{11}$B  &  $^{12}$B  &  $^{13}$B  &  $^{14}$B  \\
			\hline \vspace{-2.8mm}\\
			0 & {\color{blue} 84}  & { \color{blue} 62} & {\color{blue} 28} & {\color{blue} 5} & {\color{blue} 48}\\
			2 & { \color{blue} $ 1.5\times10^{4}$ } & { \color{blue} $ 1.6\times10^{4}$ } & { \color{blue} $ 1.2\times10^{4}$ } & { \color{blue} $ 6.0\times10^{3}$ } & {\color{blue} $ 2.8\times10^{4}$}   \\
			4 & { \color{blue} $ 5.8\times10^{5}$ } & { \color{blue} $ 8.1\times10^{5}$ } & { \color{blue} $ 8.4\times10^{5}$ } & { \color{blue} $ 6.0\times10^{5}$ } & {\color{blue} $ 2.4\times10^{6}$}\\
			6 & {\color{blue} $1.2\times10^{7}$} & {\color{blue} $2.0\times10^{7}$} & {\color{blue} $2.5\times10^{7}$} & { \color{blue} $ 2.3\times10^{7}$ } & {\color{blue} $ 8.9\times10^{7}$}\\
			8 & {\color{blue} $1.7\times10^{8}$} & {\color{blue} $3.2\times10^{8}$} & {\color{blue}$ 4.7\times10^{8}$} & $ {\color{blue}5.2\times10^{8}} $    &   $ 2.0\times 10^{9} $ \\
			10 & {\color{blue} $1.7\times10^{9}$} &  $3.7\times10^{9}$ & $6.3\times10^{9}$ & $ 8.1\times10^{9}$    &    $3.2\times10^{10} $\\
			\hline \hline
		\end{tabular} 
	\end{adjustbox}
\end{table}
We note that minima of the g.s. energy are at the same frequency for both $N_{\mbox{max}}$ = 6 and 8 for the INOY interaction. Thus, we expect to obtain the minimum at the same frequency also for $N_{\mbox{max}}$ = 10. 
Optimal frequency values for the INOY, CDB2K, N$^3$LO and N$^2$LO$_{opt}$ interactions are at $\hbar\Omega$ = 20 MeV, 14 MeV, 12 MeV and 22 MeV, respectively. Only for those values we performed the $N_{\mbox{max}}$ = 10 calculations.
We have determined the optimal frequencies for other boron isotopes as shown in Fig. \ref{basis1} corresponding to INOY and N$^2$LO$_{opt}$ interactions. Similarly, 
 we have obtained optimal frequencies for CDB2K and N$^3$LO interactions.

\begin{figure*}
	\includegraphics[width=18cm,height=8cm,clip]{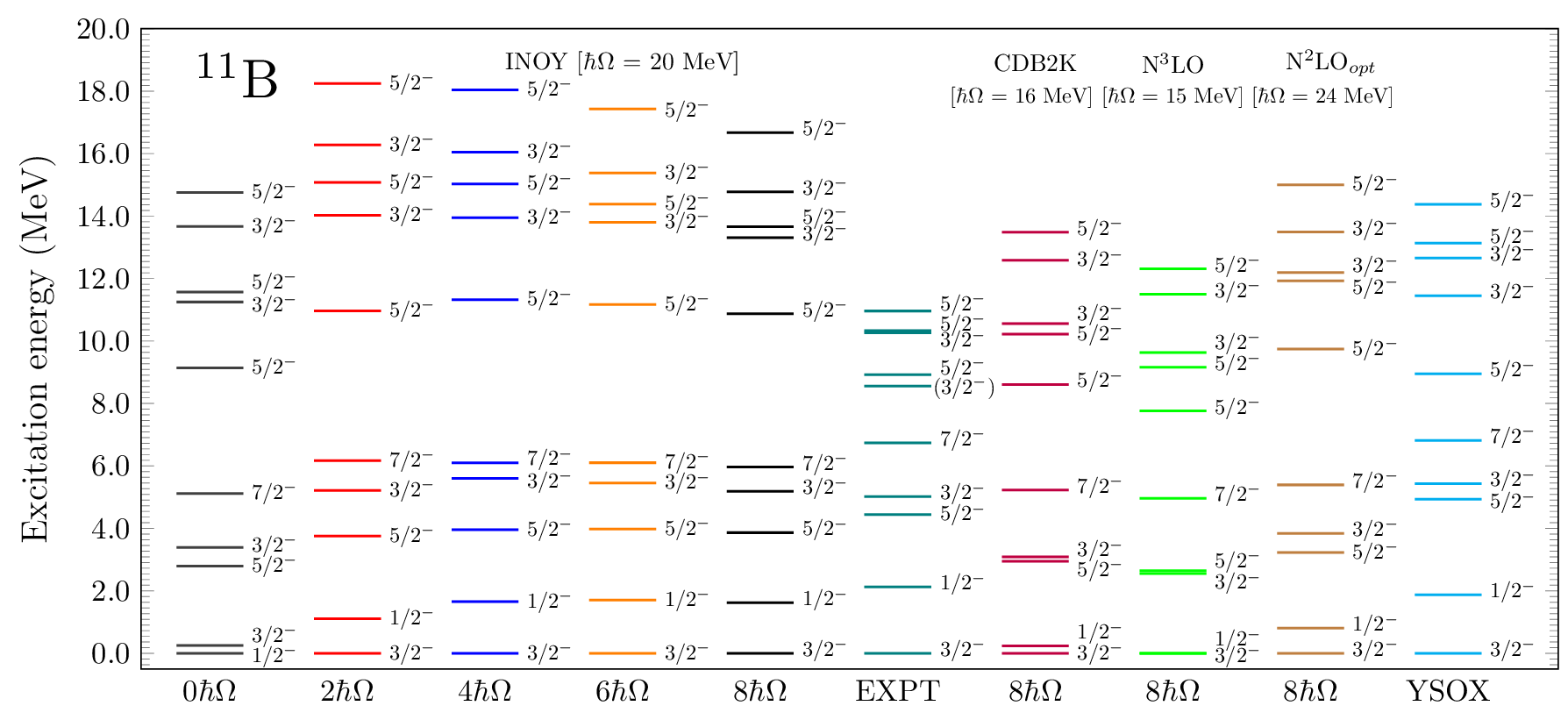}
	\includegraphics[width=18cm,height=8cm,clip]{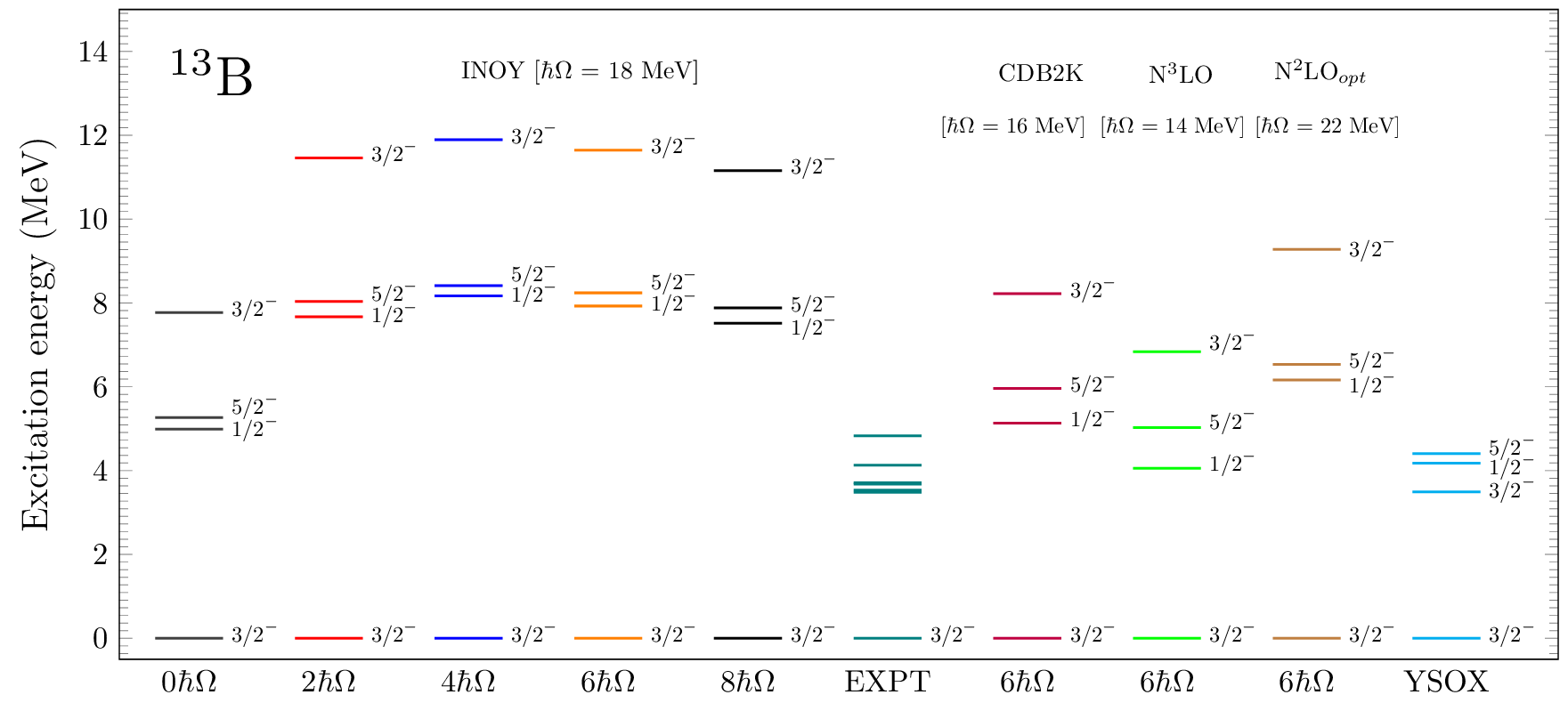}
	\caption{\label{spectra_odd} Comparison of theoretical and experimental energy spectra of $^{11,13}$B isotopes. The NCSM results are reported with the INOY, CDB2K, N$^3$LO and N2LO$_{opt}$ interactions at their optimal HO frequencies. Shell model results with the YSOX interaction is also shown.}
\end{figure*}

The NCSM results of low-lying states for boron isotopes corresponding to the INOY interaction in the basis spaces 0$\hbar\Omega$ to highest $N_{\mbox{max}}$, and for the other interactions in the highest  $N_{\mbox{max}}$ are shown in Figs. \ref{spectra_even}-\ref{spectra_odd}. From the figures, we can see how the energy states approach the experimental values.
Along with the NCSM results, we have also reported shell model results corresponding to YSOX interaction. All results are compared with experimental data. We have calculated only natural parity states for each nucleus. 

\vspace{-0.3cm}
\subsection{Energy spectra for $^{10,12,14}$B}

\begin{figure*}
	\includegraphics[width=18cm,height=8cm,clip]{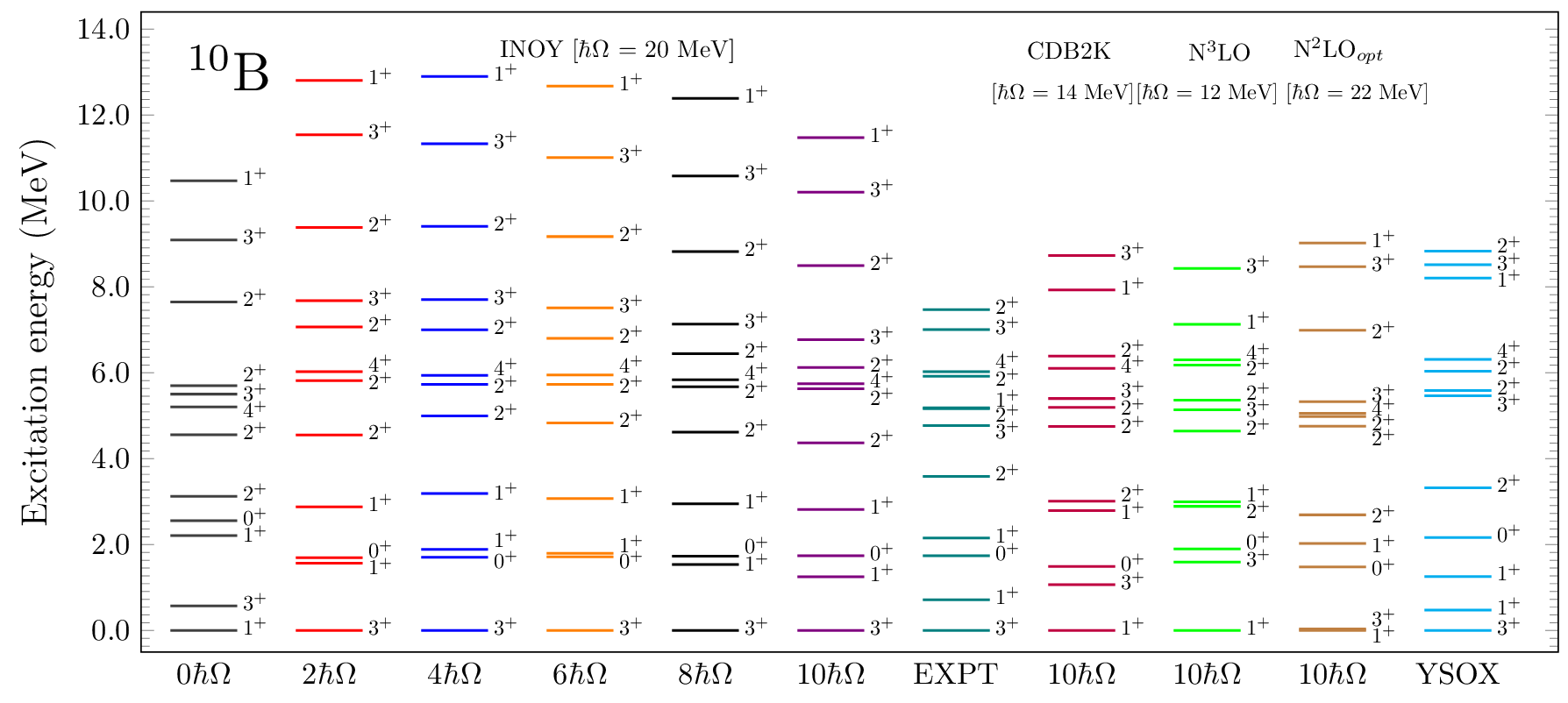}
	\includegraphics[width=18cm,height=8cm,clip]{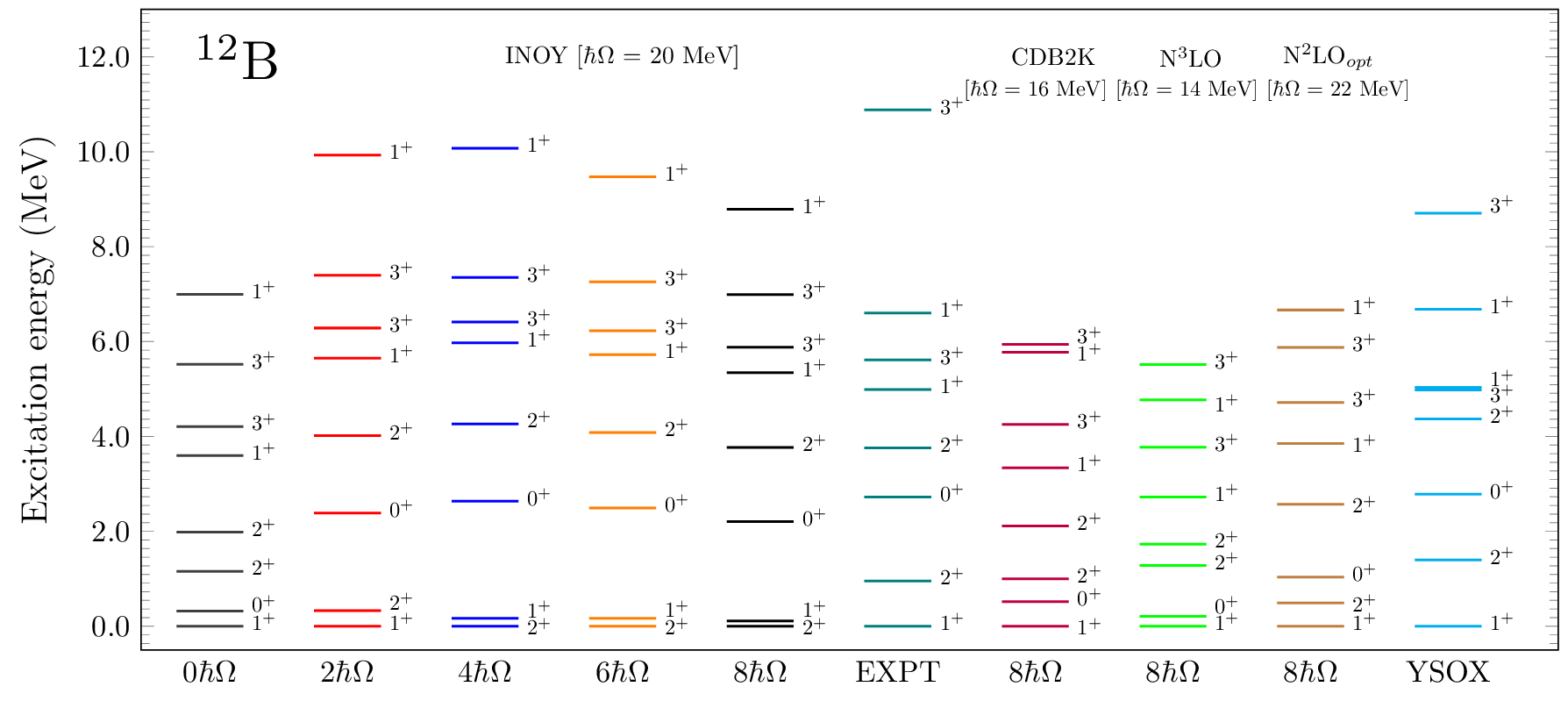}
	
	\includegraphics[width=18cm,height=8cm,clip]{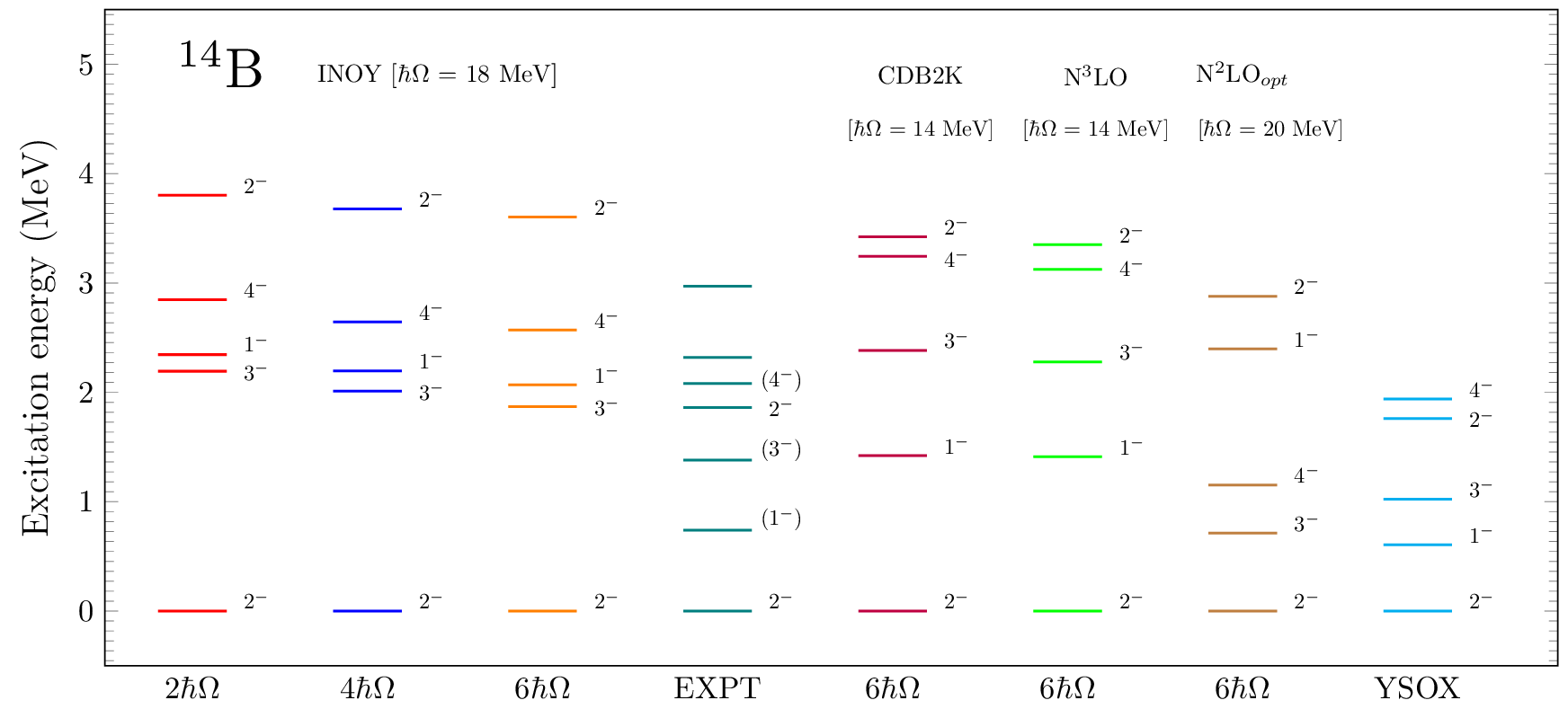}
	
	\caption{ \label{spectra_even} Comparison of theoretical and experimental energy spectra of $^{10,12,14}$B isotopes. The NCSM results are reported with the INOY, CDB2K, N$^3$LO and N2LO$_{opt}$  interactions at  their optimal HO frequencies. Shell model results with the YSOX interaction is also shown.}
	
\end{figure*}

Experimentally, the g.s. of $^{10}$B is 3$^{+}$ and the first excited state 1$^{+}$ lies 0.718 MeV above the g.s.
For the INOY interaction, we obtain the correct g.s. 3$^{+}$ as seen in the energy spectrum shown in the top panel of Fig.~\ref{spectra_even}. 
The difference between 3$^{+}$ and 1$^{+}$ states decreases as $N_{\mbox{max}}$ increases and for $N_{\mbox{max}}$ = 10, the difference is 1.250 MeV. 
Previously, the NCSM results using CDB2K interaction have been reported for $N_{\mbox{max}}$ = 8 \cite{CDB2Knocore}. In the present paper, we have extended the basis size from $N_{\mbox{max}}$ = 8 to 10 to further improve convergence. Overall, the present results are consistent with those of Ref.~\cite{CDB2Knocore}.
The CDB2K interaction is unable to reproduce the correct g.s. 3$^{+}$.
For comparison, we have also studied NCSM results with N$^{3}$LO and N$^2$LO$_{opt}$ interactions for $N_{\mbox{max}}$ = 10. These interactions predict 1$^{+}$ as the g.s. contrary to the experimental result, albeit the difference between 3$^{+}$ and 1$^{+}$ states is very small (0.035 MeV) for the N$^2$LO$_{opt}$ interaction. 
We note that the calculated 3$^{+}_{1}$ corresponding to CDB2K and N$^{3}$LO interactions is respectively, 1.069 MeV and 1.594 MeV above the 1$^{+}_{1}$ state. We can also see that the INOY interaction predicts the correct ordering of 3$^{+}$-1$^{+}$-0$^{+}$-1$^{+}$-2$^{+}$ states contrary to the phenomenological YSOX interaction.

As seen in the second panel of Fig.~\ref{spectra_even}, the INOY interaction fails to predict correct g.s. $1^{+}$ for $^{12}$B, while CDB2K, N$^{3}$LO and N$^2$LO$_{opt}$ interactions are able to predict the g.s. correctly. At the same time, it is clear that the difference between $1^{+}$ and $2^{+}$ states decreases with increasing $N_{\mbox{max}}$ for INOY interaction. So, we expect that for larger $N_{\mbox{max}}$, the g.s. would be $1^{+}$ also for the INOY interaction.
Using CDB2K and N$^{3}$LO interactions, the NCSM results are too compressed compared to experimental results. In particular, the $0^+$ state is too low.   
The N$^2$LO$_{opt}$ interaction gives the correct order of the energy levels up to 3$^{+}_{1}$ with lower energy values than the experimentally obtained energies.


For $^{14}$B, we have  reached only $N_{\mbox{max}}$ = 6 space, due to huge dimension of Hamiltonian matrix involved in the calculation.
All interactions provide the correct g.s. as $2^-$. 
Experimentally, $1^-_{1}$ and $3^-_{1}$ states are tentative, which are confirmed with the CDB2K and N$^{3}$LO interactions. These states are also confirmed with YSOX interaction. For the INOY interaction, the order of states $1^-_{1}$, $3^-_{1}$ and $2^-_{1}$, $4^-_{1}$ is reversed in comparison to the (tentative) experimental data.
The energy difference between $2^-_{1}$ and $1^-_{1}$ states is larger for all \textit{ab-initio} interactions compared to that obtained in experiment. 

\begin{table*}
	\centering
	\caption{\label{em} Electromagnetic observables of $^{10-14}$B corresponding to the largest $N_{\mbox{max}}$ at their optimal HO frequencies. Quadrupole moments, magnetic moments, g.s. energies, $E2$ and $M1$ transitions are in barn (b), nuclear magneton ($\mu_{N}$), MeV, $e^{2}$ fm$^{4}$ and $\mu_{N}^{2}$ respectively. Experimental values are taken from Refs. \cite{NNDC,Qandmag}. YSOX results are also shown for comparison.}
	\begin{tabular}{cMMMMMM}
		\hline
		\hline \vspace{-2.8mm}\\
		$^{10}$B & EXPT & INOY & CDB2K & N$^3$LO & N$^2$LO$_{opt}$ & YSOX \\
		\hline \vspace{-2.8mm}\\
		Q($3^{+}$) & 0.0845(2) & 0.061 & 0.071 & 0.077 & 0.067 &0.073 \\
		$\mu$($3^{+}$) & 1.8004636(8) &1.836  &1.852  &1.856& 1.838  &1.806 \\
		E$_{g.s.}$($3^{+}$)& -64.751 & -63.433 &-54.979&-53.225 & -54.181 & -65.144 \\
		$B(E2;3_{1}^{+}$ $\rightarrow$ $1_{1}^{+}$) & 1.777(9)&0.911 & 2.091 &2.686 & 1.482 &0.757 \\
		$B(M1;2_{1}^{+}$ $\rightarrow$ $3_{1}^{+}$) & 0.00047(27) & 0.0007 &0.002 &0.003 &0.0001 &0.004\\
		\hline \vspace{-2.8mm}\\
		$^{11}$B & EXPT & INOY & CDB2K & N$^3$LO & N$^2$LO$_{opt}$ & YSOX\\
		\hline \vspace{-2.8mm}\\
		Q($3/2^{-}$) & 0.04059(10) & 0.027 & 0.030 & 0.031 & 0.029&0.043  \\
		$\mu$($3/2^{-}$) & 2.688378(1) & 2.371 & 2.537 & 2.622 & 2.366 &2.501\\
		E$_{g.s.}$($3/2^{-}$) & -76.205& -74.926&-66.034&-62.915&-59.993&-76.686\\
		$B(E2;7/2_{1}^{-}$ $\rightarrow$ $3/2_{1}^{-}$) &1.83(44) &0.814 &1.258 &1.478 & 1.032 & 3.118\\
		$B(M1;3/2_{1}^{-}$ $\rightarrow$ $1/2_{1}^{-}$) & 0.519(18) & 0.708 & 0.976 & 1.051 & 0.766&0.835\\
		\hline \vspace{-2.8mm}\\
		$^{12}$B & EXPT & INOY & CDB2K & N$^3$LO & N$^2$LO$_{opt}$ & YSOX\\
		\hline \vspace{-2.8mm}\\
		Q($1^{+}$) & 0.0132(3) & 0.009 & 0.009 & 0.010 & 0.010& 0.014\\
		$\mu$($1^{+}$) & 1.003(1) &0.561 & 0.134 & 0.022& 0.282&0.737\\
		E$_{g.s.}$($1^{+}$)& -79.575&-78.304&-69.350&-68.062&-61.226&-79.264\\
		$B(M1;1_{1}^{+}$ $\rightarrow$ $0_{1}^{+}$) &NA  & 0.047 & 0.078& 0.086 & 0.066& 0.026\\
		$B(M1;2_{1}^{+}$ $\rightarrow$ $1_{1}^{+}$) & 0.251(36) & 0.125 &0.197 & 0.339 & 0.170&0.204\\
		\hline \vspace{-2.8mm}\\
		$^{13}$B & EXPT & INOY & CDB2K & N$^3$LO & N$^2$LO$_{opt}$& YSOX \\
		\hline \vspace{-2.8mm}\\
		Q($3/2^{-}$) & 0.0365(8) & 0.025 & 0.029 & 0.031 & 0.028 &  0.042 \\
		$\mu$($3/2^{-}$) & 3.1778(5) & 2.844 & 2.815 & 2.830 & 2.781 &2.959 \\
		E$_{g.s.}$($3/2^{-}$)&-84.454 & -85.205 & -75.856& -74.716 & -65.624 & -84.185  \\
		$B(E2;5/2_{1}^{-}$ $\rightarrow$ $1/2_{1}^{-}$) &NA& 1.800 & 2.281 & 2.721 & 1.990 &0.787  \\
		$B(M1;3/2_{1}^{-}$ $\rightarrow$ $1/2_{1}^{-}$) & NA &  0.984 & 1.035 & 1.065 & 0.982 & 0.729\\
		\hline \vspace{-2.8mm}\\
		$^{14}$B & EXPT & INOY & CDB2K & N$^3$LO & N$^2$LO$_{opt}$ & YSOX\\
		\hline \vspace{-2.8mm}\\
		Q($2^{-}$) & 0.0297(8) & 0.016 & 0.025 & 0.025 & 0.004 &0.026\\
		$\mu$($2^{-}$) & 1.185(5) & 0.778 &  0.926 & 0.914 &0.550&0.614 \\
		E$_{g.s.}$($2^{-}$)&-85.422&-82.002&-76.929&-77.549&-51.413&-84.454 \\
		$B(M1;2_{1}^{-}$ $\rightarrow$ $1_{1}^{-}$) & NA& 2.579 & 2.457 & 2.436 &2.755 &2.656 \\
		\hline
                \hline
	\end{tabular}
\end{table*}
\subsection{Energy spectra for $^{11,13}$B}
\vspace{-0.1cm}
For $^{11}$B, we employed HO frequencies of 20 MeV, 16 MeV and 24 MeV for the INOY, CDB2K and N$^2$LO$_{opt}$ interaction, respectively. For N$^3$LO interaction, optimal frequency is taken to be 15 MeV from Ref.~\cite{10B_PRL}.
The $3/2^{-}$ state is the experimental g.s. of $^{11}$B. Our NCSM calculations reproduce the correct g.s. with all four interactions. We get correct excited states up to $\sim$ 7 MeV  with all interactions except the N$^3$LO.
The experimental g.s. energy of the $3/2^{-}$ state is -76.205 MeV. With the INOY interaction, we obtain the energy of -74.9 MeV for  this state, fairly close to the experimental value. 
For N$^3$LO interaction, $3/2^{-}$ and $1/2^{-}$ states are almost degenerate, while the INOY gives a splitting close to experimental. This splitting depends on the strength of the spin-orbit interaction, which is apparently the largest for the INOY interaction.
We note that the energy gap between the states $7/2^{-}_{\,1}$ and $5/2^{-}_{\,2}$ obtained using the INOY interaction is very large compared to the experimental value. This could be because the optimal HO frequency is chosen with respect to the g.s. which is then used to predict the whole energy spectrum. It is possible that a faster convergence of the excited states could be achieved with a different optimal frequency. 

\begin{figure*}
	\includegraphics[width=8cm,height=7.3cm]{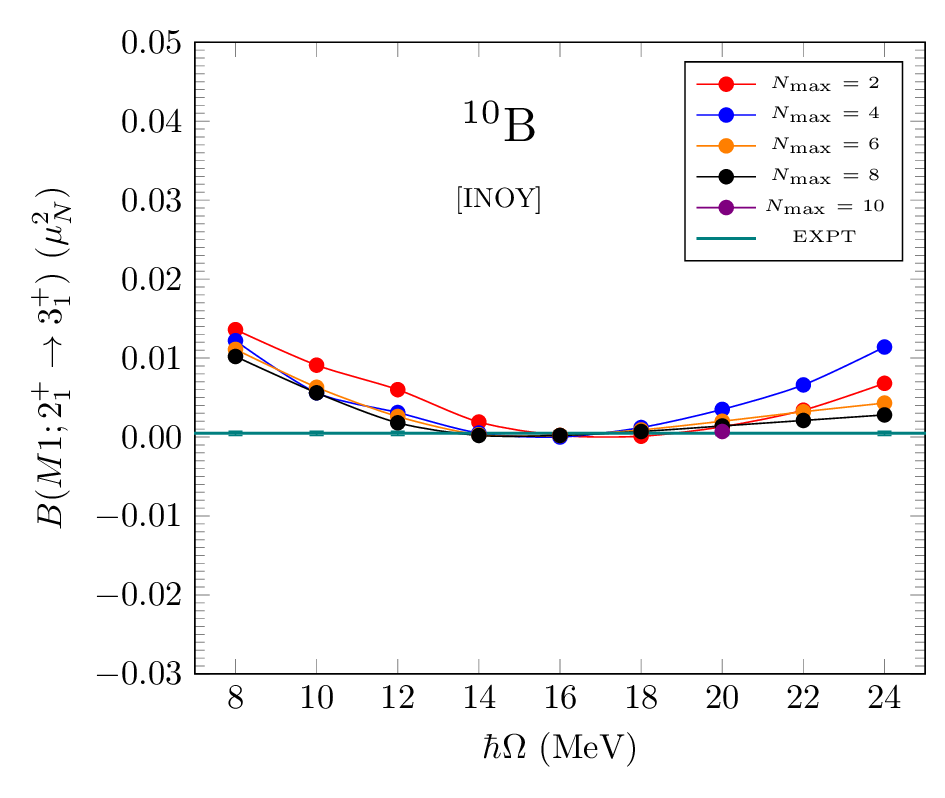}
	\includegraphics[width=8cm]{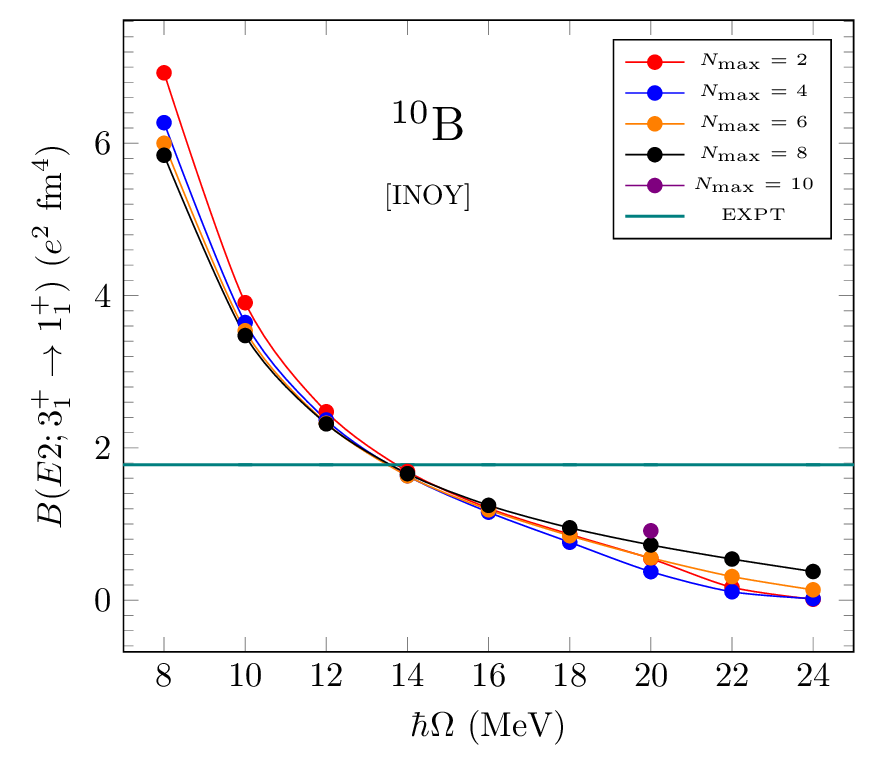}
	\includegraphics[width=8cm,height=7.3cm]{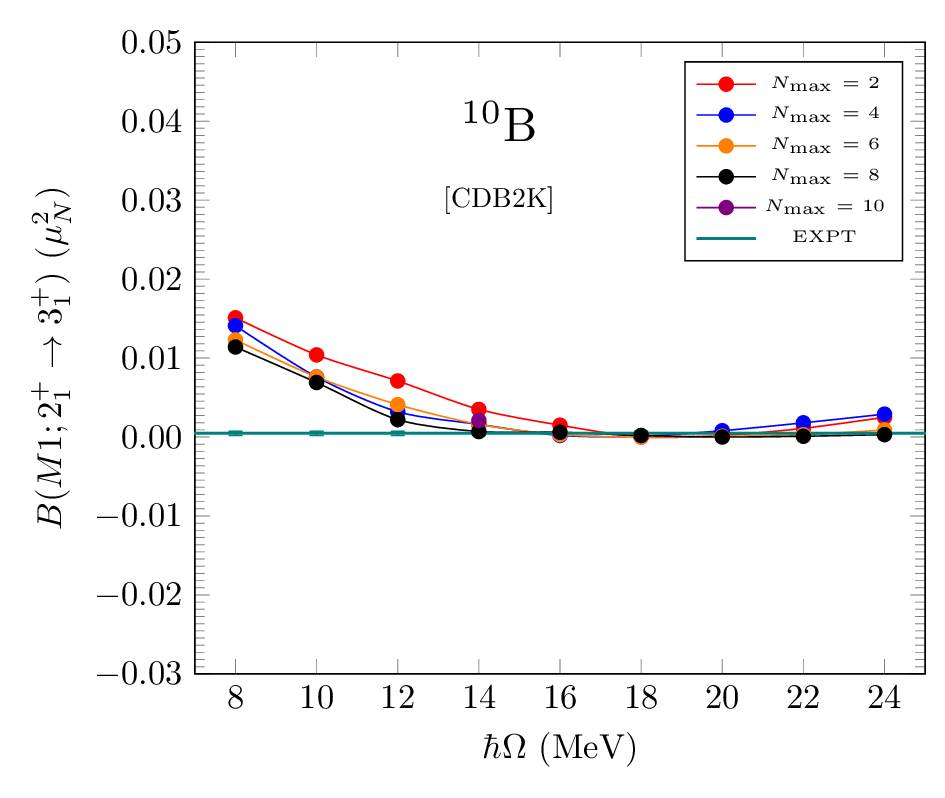}
	\includegraphics[width=8cm]{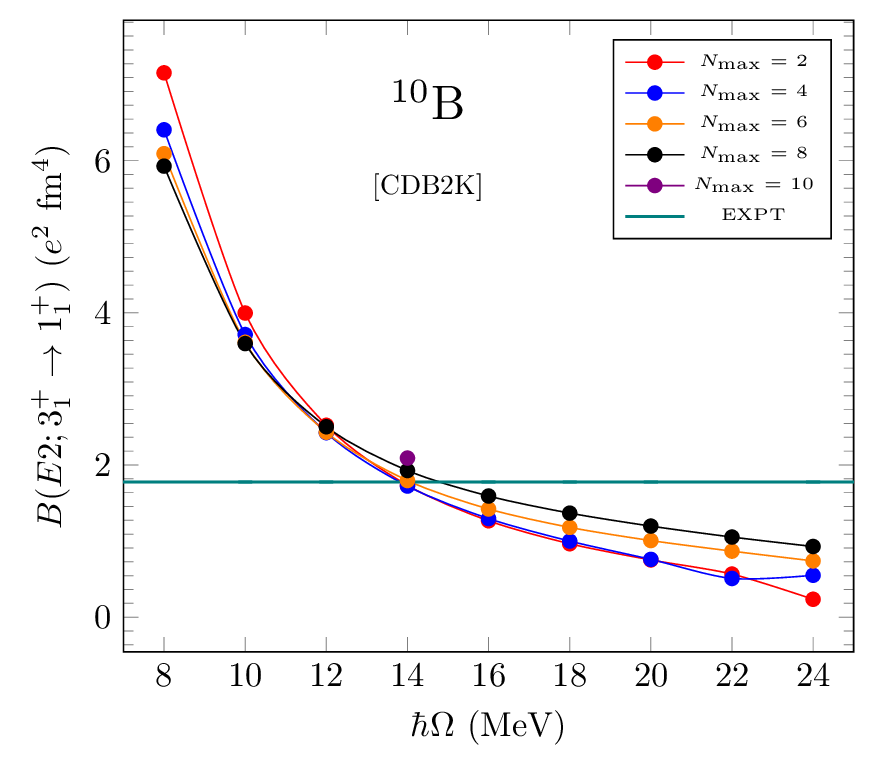}
	\includegraphics[width=8cm,height=7.3cm]{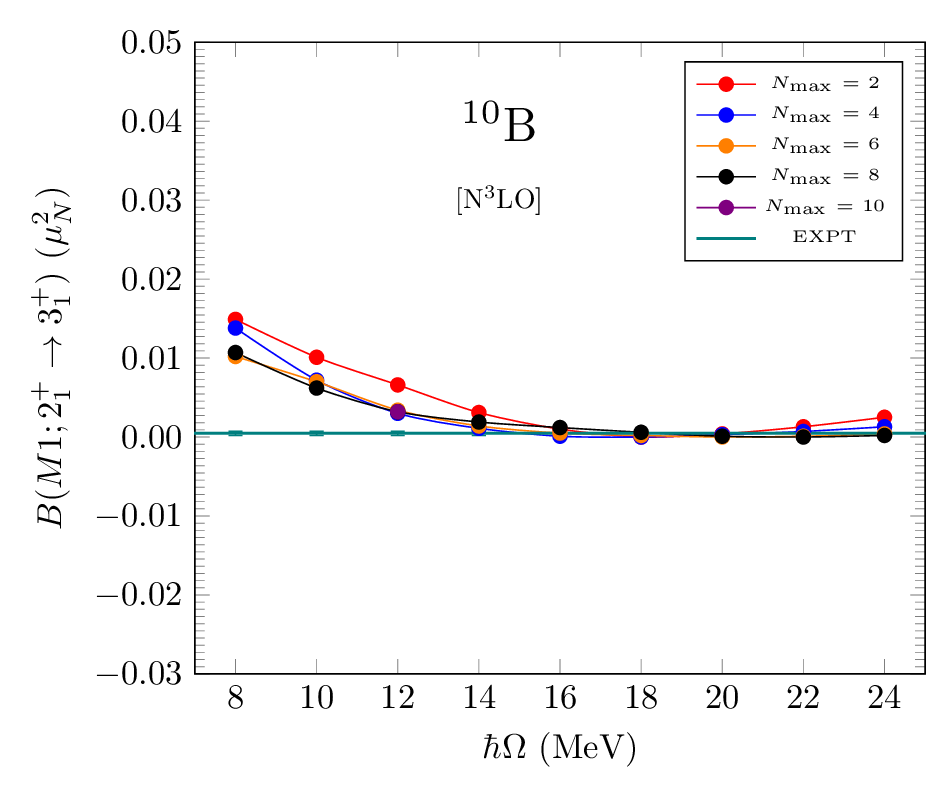}
	\includegraphics[width=8cm]{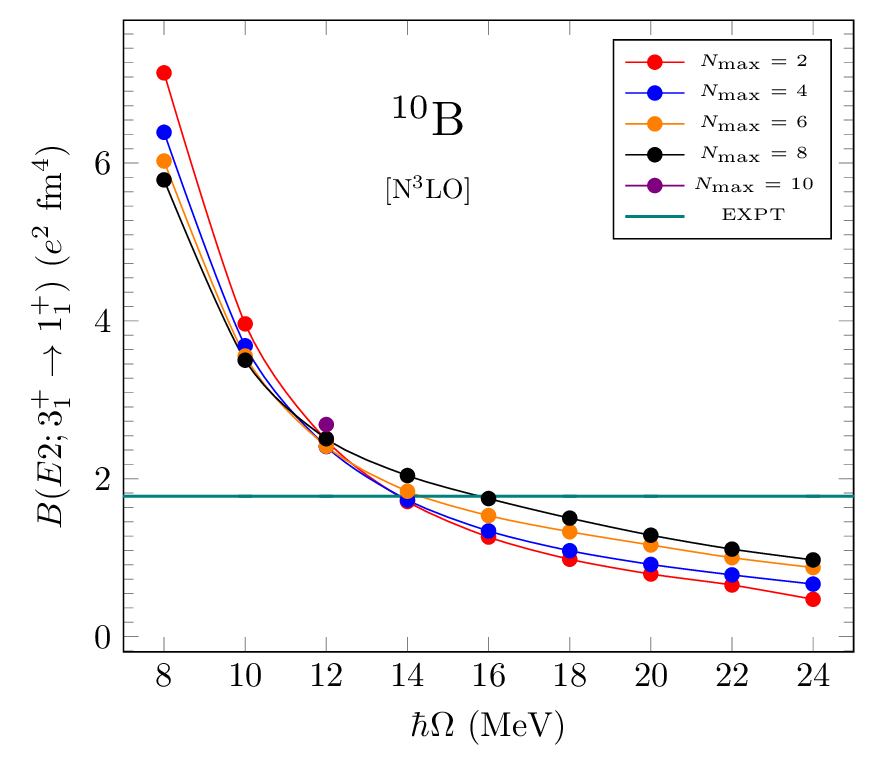}
	
	\caption{ \label{EM} Variation of B(M1:$2_1^+ \rightarrow 3_1^+$) and B(E2:$3_1^+ \rightarrow 1_1^+$) for $^{10}$B with HO frequency for $N_{\mbox{max}}$ = 2 to 10, corresponding to the INOY, N$^3$LO and CDB2K interactions. Experimental values are shown by horizontal line with  uncertainty. }
\end{figure*}
Our NCSM calculations have been performed up to $N_{\mbox{max}}$ = 8 for $^{13}$B, for which
	we obtain correct g.s. with all interactions.  The energy difference between theoretical and experimental excited states is rather large, which makes it difficult to use the present calculations for assigning experimentally unknown spin and parity to the excited states. 

\section{Electromagnetic properties}
\vspace{-0.05cm}
Table \ref{em} contains quadrupole moments ($Q$), magnetic moments ($\mu$), g.s. energies (E$_{g.s.}$), reduced electric quadrupole transition probabilities ($B(E2)$) and reduced magnetic dipole transition probabilities ($B(M1)$). 
Only one-body electromagnetic operators were considered.
The experimental binding energy of $^{10}$B is -64.751 MeV. The INOY interaction underbinds the $^{10}$B nucleus by 1.32 MeV while YSOX interaction  overbinds this by 0.39 MeV. The other used realistic interactions underestimate the experimental binding energy more significantly.
The g.s. $Q$ and $\mu$ moments of $^{10,11}$B are in a reasonable agreement with experiment for all interactions.
On the other hand, the calculated $B(E2;3_1^+\rightarrow 1_1^+)$ value for $^{10}$B varies substantially. Similarly, we find interaction dependence and stronger disagreements with experiment for the $^{12,13,14}$B g.s. moments. 
We predict several $B(E2)$ and $B(M1)$ values for $^{12-14}$B which are not yet measured experimentally.
\begin{figure*}
	\includegraphics[width=8.8cm,height=7cm,clip]{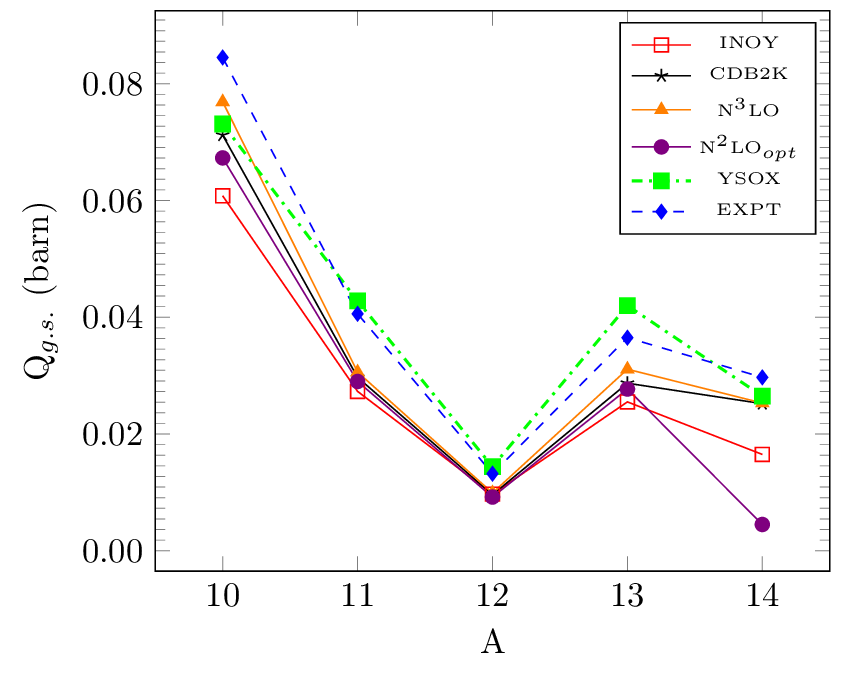}
	\includegraphics[width=8.8cm,height=7cm,clip]{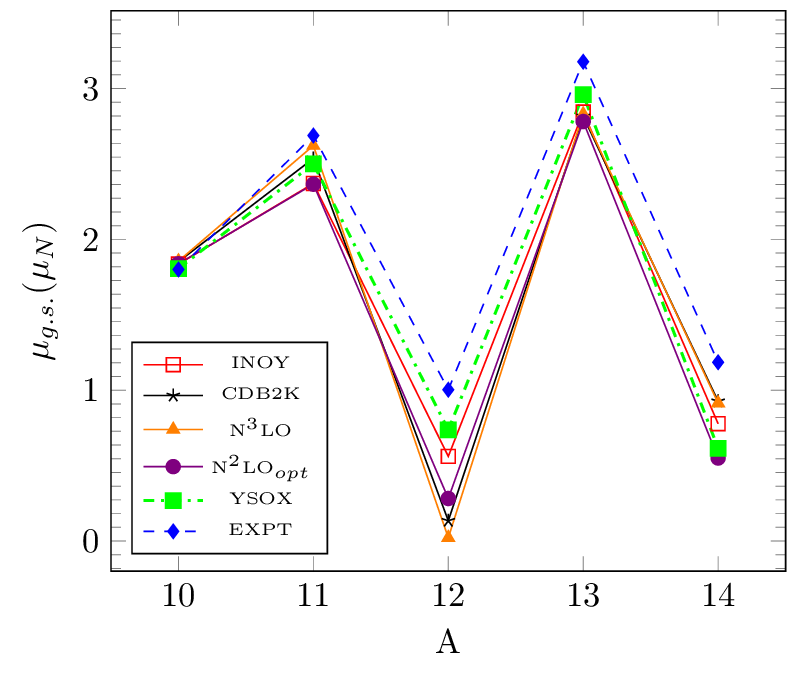}
	\caption{Ground state quadrupole and magnetic moment dependencies on the mass number of the studied boron isotopes. NCSM results obtained at the largest accessible $N_{\rm max}$ space with the optimal frequency are shown. Experimental values are taken from Ref. \cite{Qandmag}.}
	\label{basis2}
\end{figure*}
In  Fig. \ref{EM}, we show $B(M1; 2_1^+ \rightarrow 3_1^+)$ and $B(E2; 3_1^+ \rightarrow 1_1^+)$ transition strengths corresponding to 
different $N_{\mbox{max}}$ and  $\hbar\Omega$ for $^{10}$B with the INOY, CDB2K and N$^{3}$LO interactions. $B(M1; 2_1^+\rightarrow 3_1^+)$ curves become flat,  which means they become independent of $N_{\mbox{max}}$ and  $\hbar\Omega$. So, the convergence of the $B(M1)$ result is obtained at smaller $\hbar\Omega$ and lower $N_{\mbox{max}}$. 
As discussed, e.g., in Refs. \cite{stetcu1,stetcu2}, it is a big task to compute the $E2$ transition operator, as it depends on the long-range correlations in the nucleus i.e. the tails of nuclear wave functions.
From  Fig. \ref{EM}, we can see that $B(E2)$ value  varies even for large value of the $N_{\mbox{max}}$ parameter. The best $B(E2)$ value is then taken where these curves become flat, although clearly we have not reached convergence within the model spaces used in this work. 

The quadrupole and magnetic moments of the studied isotopes are summarized in Fig.~\ref{basis2}. Overall, the experimental trends are well reproduced for both observables although the NCSM calculations systematically under predict the experimental quadrupole moments.

\begin{figure}
	\includegraphics[width=8.8cm,height=7cm,clip]{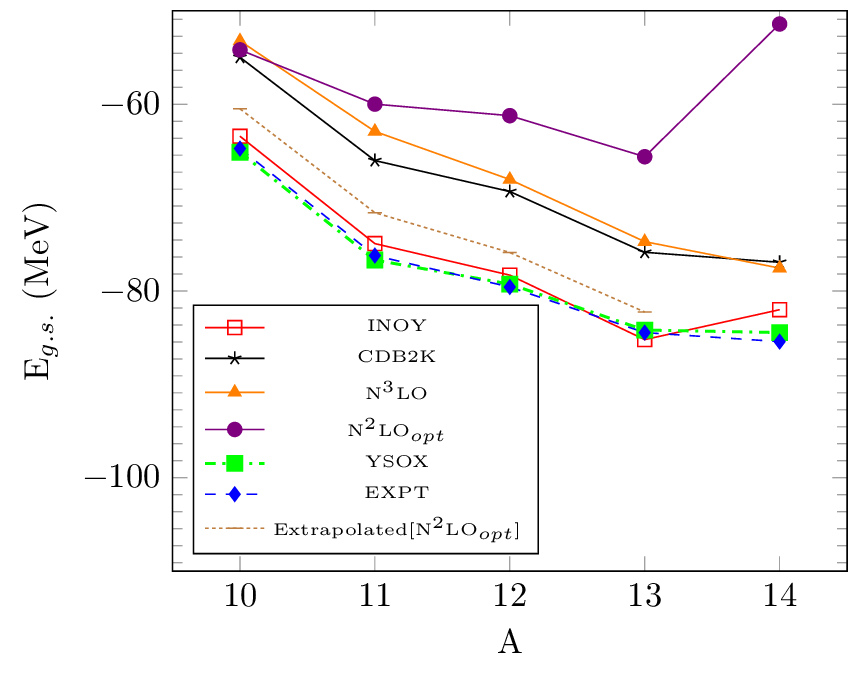}
	\caption{\label{energy_dependence_all} Dependence of the calculated g.s. energies on A of boron isotopes with INOY, CDB2K, N$^{3}$LO, N$^2$LO$_{opt}$, YSOX interactions and compared with experimental energies. NCSM results obtained at the largest accessible $N_{\rm max}$ space with the optimal frequency are shown.}
\end{figure}

In Fig. \ref{energy_dependence_all}, the dependence of the calculated g.s. energies on the mass number of boron isotopes is plotted with INOY, CDB2K, N$^{3}$LO, N$^2$LO$_{opt}$, YSOX interactions and compared with experimental energies. NCSM results obtained at the largest accessible $N_{\rm max}$ space with the optimal frequency are shown. From Fig. \ref{energy_dependence_all}, we can conclude that INOY interaction provides better description for g.s. energy than other used {\it ab initio} interactions.

For the N$^2$LO$_{opt}$ interaction, we have extrapolated the g.s. energy using an exponential fitting function $E_{g.s.}(N_{max})= a~exp(-bN_{max}) + E_{g.s.}(\infty)$ with 
 $E_{g.s.}(\infty)$ the value of g.s. energy at $N_{max}$ $\rightarrow$ $\infty$. In particular, we have used last three $N_{max}$ points in the extrapolation  procedure. For $^{14}$B, no meaningful extrapolation was possible.

\begin{figure*}
	\includegraphics[width=8cm]{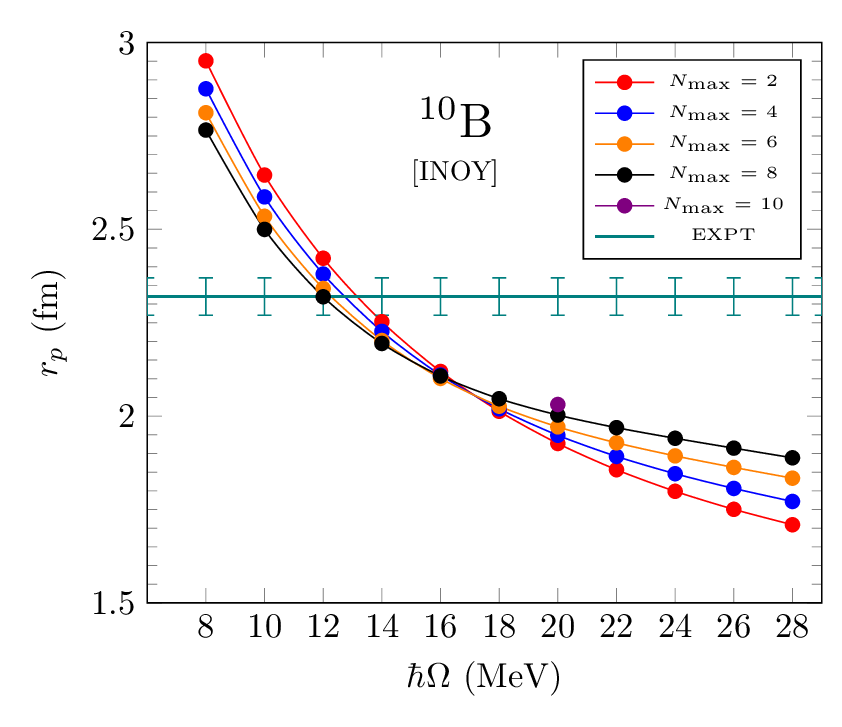}
	\includegraphics[width=8cm]{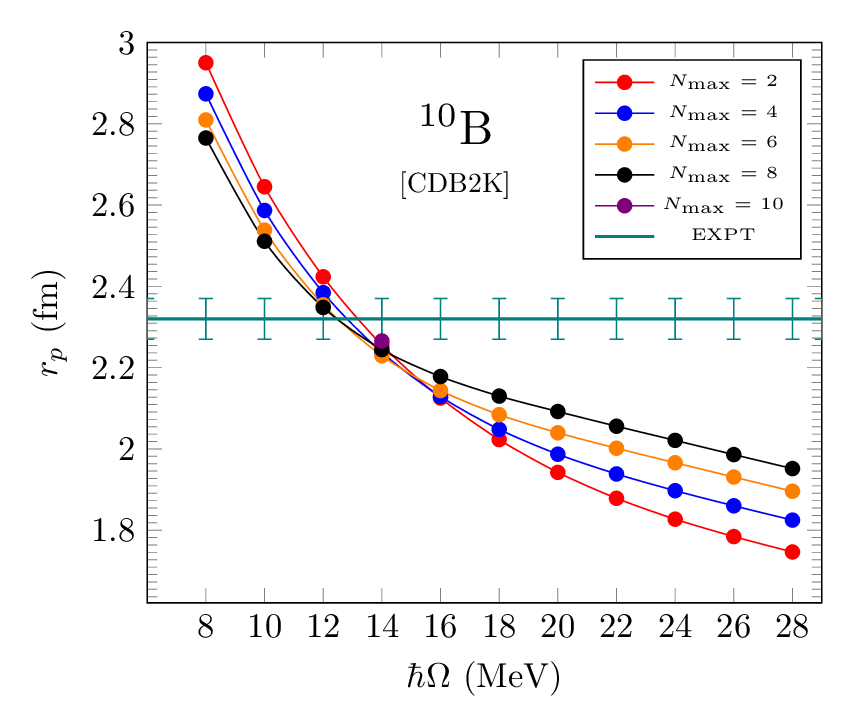}
	\includegraphics[width=8cm]{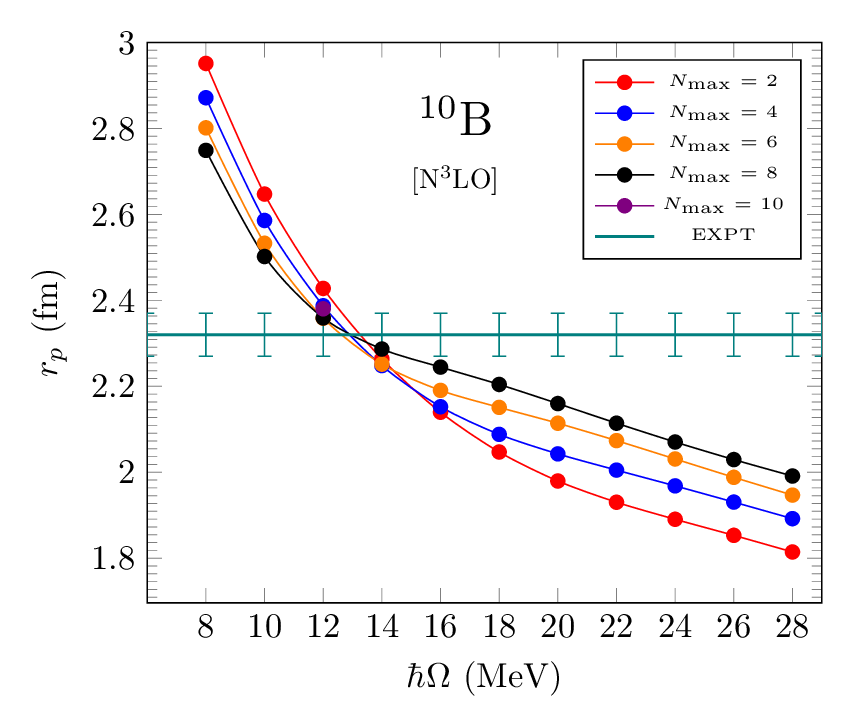}
	
	\caption{ \label{radii10B} Variation of $r_{p}$ of $^{10}$B with HO frequency for $N_{\mbox{max}}$ = 2 to 10, corresponding to the INOY, N$^3$LO and CDB2K interactions.
	The horizontal line shows the experimental value with the vertical bars representing uncertainty.}	 
\end{figure*}

\begin{figure*}
	\includegraphics[width=8cm]{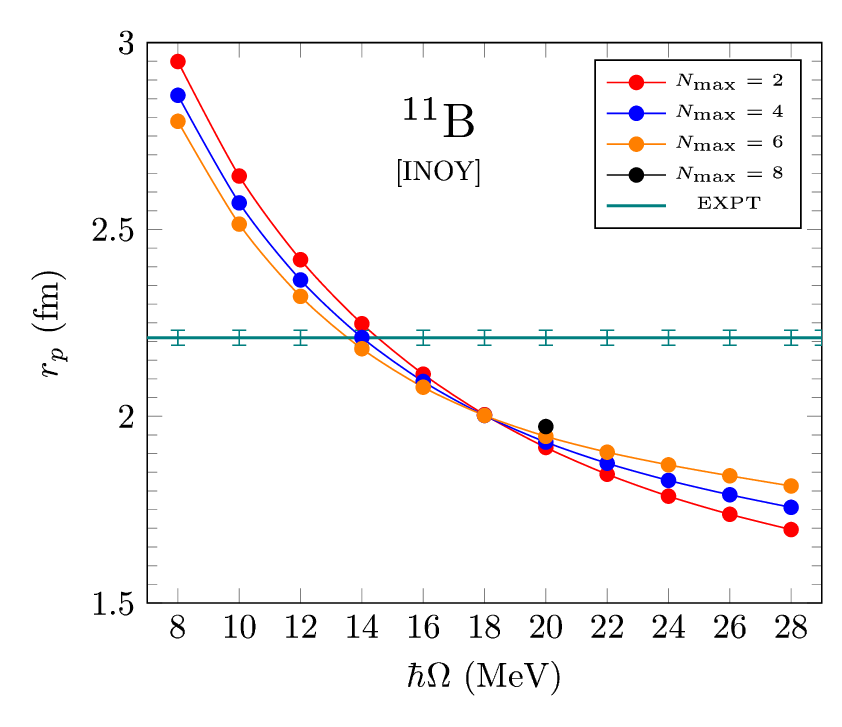}
	\includegraphics[width=8cm]{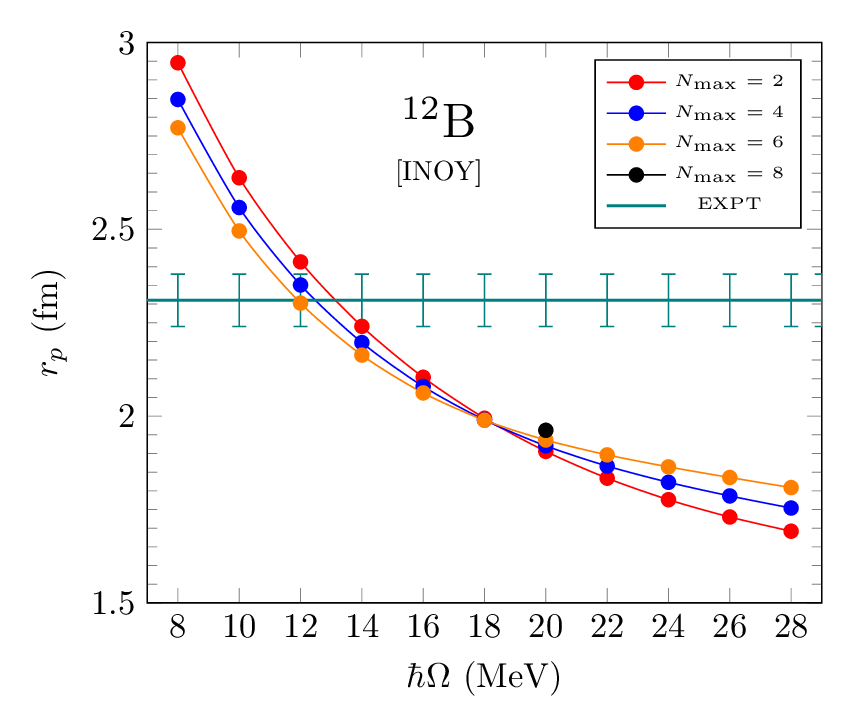}	
          \includegraphics[width=8cm]{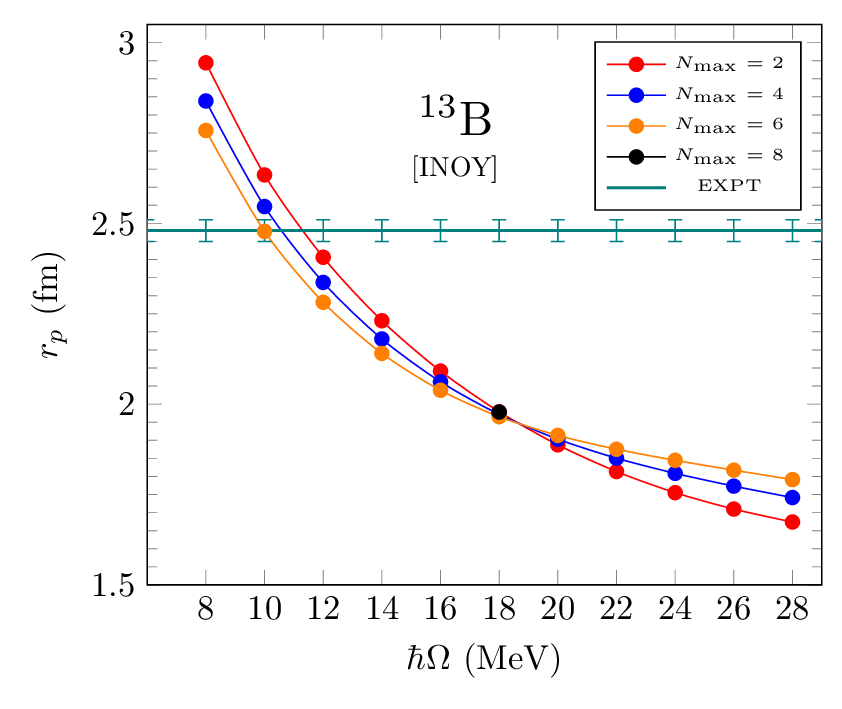}
	\includegraphics[width=8cm]{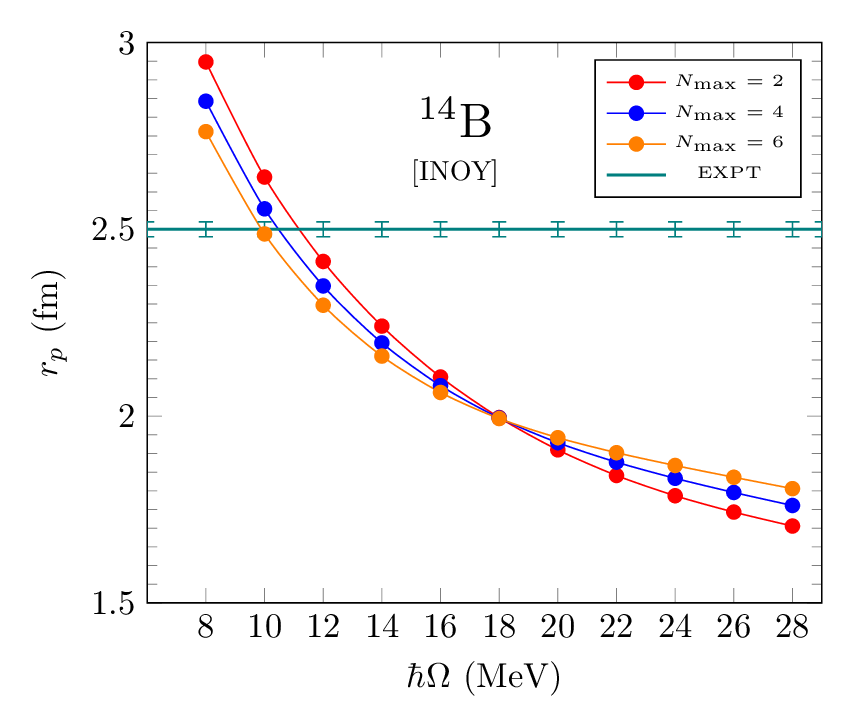}
	\caption{ \label{radiiB} Variation of $r_{p}$ of $^{11,12,13,14}$B with HO frequency for different $N_{\mbox{max}}$, corresponding to the INOY interaction.
	The horizontal line shows the experimental value with the vertical bars representing uncertainty.}	 
\end{figure*}
\section{Point-proton radii}

In Table \ref{radii}, we have presented point-proton radii (\textit{r}$_{p}$) using NCSM with INOY, CDB2K and N$^3$LO interactions at their optimal frequencies along with experimentally observed radii \cite{chargechanging}. The INOY interaction considerably underestimates the radii.  For $^{10,11}$B, the CDB2K and N$^3$LO interactions produce better results, with the former slightly underestimating and the latter slightly overestimating the radii. For $^{12-14}$B, the radii are underestimated for all interactions.

\begin{table}[h]
	\centering
	\caption{\label{radii} Calculated point-proton radii ($r_{p}$) of $^{10-14}$B with INOY, CDB2K and N$^3$LO interactions at highest $N_{\mbox{max}}$ corresponding  to their optimal HO frequencies. Experimental point-proton radii are taken from Ref. \cite{chargechanging}. The point-proton radii are given in fm.}
	\begin{tabular}{cMMMM}
		\hline
		\hline \vspace{-2.8mm}\\
		$ r_{p}$ & EXPT & INOY & CDB2K & N$^3$LO  \\
		\hline \vspace{-2.8mm}\\
		$^{10}$B &2.32(5) & 2.03& 2.27 & 2.38 \\
		
		$^{11}$B & 2.21(2) & 1.97 & 2.15& 2.24 \\
		
		$^{12}$B &2.31(7) & 1.96& 2.13 & 2.23  \\
		
		$^{13}$B &2.48(3) & 1.98 & 2.10 & 2.20  \\
		
		$^{14}$B & 2.50(2)&1.99 & 2.18 & 2.20  \\
		\hline \hline
	\end{tabular}
\end{table}

In Fig. \ref{radii10B},  we present the variation of $^{10}$B $r_p$ with frequency and $N_{\mbox{max}}$ for INOY, CDB2K and N$^3$LO interaction.  With the enlargement of basis size $N_{\mbox{max}}$,  the
dependence of $r_{p}$ on frequencies decreases. 
The curves of $r_p$ corresponding to different $N_{\mbox{max}}$ intersect each-other approximately at the same point. We take this crossing point as an estimate of the converged radius
\cite{PLB,Phys.Rev.C90034305(2014)}. 
In particular, we consider the intersection point of the curves at  the highest successive $N_{\mbox{max}}$ as an estimate of the converged radius. 
In this way, we obtain $^{10}$B point-proton radii for INOY, CDB2K and N$^3$LO interactions 2.14, 2.30 and 2.36 fm, respectively.

Similarly, we have shown variation of $r_{p}$ with frequency and $N_{\mbox{max}}$ for other isotopes corresponding to INOY interaction in Fig. \ref{radiiB}. Obtained 
	 $r_{p}$ values for $^{11}$B,$^{12}$B,$^{13}$B and $^{14}$B are 2.00, 1.99, 1.95 and 1.99 fm, respectively. However, even with this determination of the radii, the experimental trend is not reproduced.

We can conclude that the CDB2K and N$^3$LO interactions give radii which are much closer to experimental value than the radii obtained with the INOY interaction. To some extent this is not surprising given the fact that those interactions underbind the studied isotopes.
We have obtained different optimal  frequencies for the energy spectra and the point-proton radii. Similar findings were reported for $^{12}$C using Daejeon16 and JISP16 interactions in Ref. \cite{PLB}.

\section{Conclusions}

In this work, we have applied \textit{ab-initio} no-core shell model to obtain spectroscopic properties of boron isotopes using INOY, CDB2K,  N$^{3}$LO and N$^{2}$LO$_{opt}$ nucleon-nucleon interactions.
We have calculated low-lying spectra and other observables with all four interactions and, in addition, compared the NCSM results with shell model using YSOX valence-space effective interaction. 
We were able to correctly reproduce the g.s. spin of $^{10}$B only with the INOY \textit{NN} interaction. Overall, the INOY interaction reproduced quite reasonably g.s. energies of all the studied isotopes, $^{10-14}$B.

Considering electromagnetic properties, we have obtained fast convergence for $M1$ values, whereas, converging $E2$ observables is a computational challenge. The INOY interaction again appears to do better than the other interactions in the reproduction of the $M1$ observables for all isotopes. 


Concerning proton radii, we find that optimal frequency obtained from the minima of the g.s. energy curves and that obtained from the intersection of radii curves could be different. In this case, the CDB2K and N$^3$LO interactions give radii which are much closer to experimental value than the radii obtained with the INOY interaction.

The present study confirms that non-locality in the \textit{NN} interaction can account for some of the many-nucleon force effects. The non-local \textit{NN} interaction like INOY can provide a quite reasonable description of ground-state energies, excitation spectra and selected electromagnetic properties, e.g., magnetic moments and $M1$ transitions. However, the description of nuclear radii and consequently of the density remains unsatisfactory. Recent studies show that the inclusion of the 3\textit{N} interaction, in particular 3\textit{N} interaction with non-local regulators, is essential for a correct simultaneous description of nuclear binding and nuclear size~\cite{Nmax10,N2LOsat,NN3Nlnl}.  


\vspace*{-0.1cm}
\section*{ACKNOWLEDGMENTS}

We would like to thank Christian Forss\'en for making available the pAntoine code.
We thank Toshio Suzuki for the YSOX interaction.
P.C. acknowledges financial support from MHRD (Government of India) for her PhD thesis work. P.C.S. acknowledges a research grant from SERB (India), CRG/2019/000556. P.N. acknowledges support from the NSERC Grant No. SAPIN-2016-00033. TRIUMF receives federal funding via a contribution agreement with the National Research Council of Canada.


\bibliographystyle{utphys}
  \bibliography{references}

\end{document}